\documentclass[a4paper,11pt]{article}
\usepackage{jcappub} 
\usepackage{natbib}
\usepackage{ulem}
\usepackage[T1]{fontenc}
\usepackage{slashed}
\usepackage{physics}
\usepackage{xcolor}
\usepackage{overpic}
\usepackage{bm}
\usepackage{overpic}
\usepackage{subcaption}
\usepackage{comment}
\definecolor{seagreen}{rgb}{0.05, 0.65, 0.20}
\definecolor{darkgreen}{HTML}{1B5E20}

\newcommand{\che}[1]{{\color{seagreen}\bf[CE:  {#1}]}}

\newcommand\aj{\ref@jnl{AJ}}
\newcommand\psj{\ref@jnl{PSJ}}
\newcommand\araa{\ref@jnl{ARA\&A}}
\newcommand\apj{\ref@jnl{ApJ}}
\newcommand\apjl{\ref@jnl{ApJL}}     
\newcommand\apjs{\ref@jnl{ApJS}}
\newcommand\ao{\ref@jnl{ApOpt}}
\newcommand\apss{\ref@jnl{Ap\&SS}}
\newcommand\aap{\ref@jnl{A\&A}}
\newcommand\aapr{\ref@jnl{A\&A~Rv}}
\newcommand\aaps{\ref@jnl{A\&AS}}
\newcommand\azh{\ref@jnl{AZh}}
\newcommand\baas{\ref@jnl{BAAS}}
\newcommand\icarus{\ref@jnl{Icarus}}
\newcommand\jaavso{\ref@jnl{JAAVSO}}  
\newcommand\jrasc{\ref@jnl{JRASC}}
\newcommand\memras{\ref@jnl{MmRAS}}
\newcommand\mnras{\ref@jnl{MNRAS}}
\newcommand\pra{\ref@jnl{PhRvA}}
\newcommand\prb{\ref@jnl{PhRvB}}
\newcommand\prc{\ref@jnl{PhRvC}}
\newcommand\prd{\ref@jnl{PhRvD}}
\newcommand\pre{\ref@jnl{PhRvE}}
\newcommand\prl{\ref@jnl{PhRvL}}
\newcommand\pasp{\ref@jnl{PASP}}
\newcommand\pasj{\ref@jnl{PASJ}}
\newcommand\qjras{\ref@jnl{QJRAS}}
\newcommand\skytel{\ref@jnl{S\&T}}
\newcommand\solphys{\ref@jnl{SoPh}}
\newcommand\sovast{\ref@jnl{Soviet~Ast.}}
\newcommand\ssr{\ref@jnl{SSRv}}
\newcommand\zap{\ref@jnl{ZA}}
\newcommand\nat{\ref@jnl{Nature}}
\newcommand\iaucirc{\ref@jnl{IAUC}}
\newcommand\aplett{\ref@jnl{Astrophys.~Lett.}}
\newcommand\apspr{\ref@jnl{Astrophys.~Space~Phys.~Res.}}
\newcommand\bain{\ref@jnl{BAN}}
\newcommand\fcp{\ref@jnl{FCPh}}
\newcommand\gca{\ref@jnl{GeoCoA}}
\newcommand\grl{\ref@jnl{Geophys.~Res.~Lett.}}
\newcommand\jcp{\ref@jnl{JChPh}}
\newcommand\jgr{\ref@jnl{J.~Geophys.~Res.}}
\newcommand\jqsrt{\ref@jnl{JQSRT}}
\newcommand\memsai{\ref@jnl{MmSAI}}
\newcommand\nphysa{\ref@jnl{NuPhA}}
\newcommand\physrep{\ref@jnl{PhR}}
\newcommand\physscr{\ref@jnl{PhyS}}
\newcommand\planss{\ref@jnl{Planet.~Space~Sci.}}
\newcommand\procspie{\ref@jnl{Proc.~SPIE}}

\newcommand\actaa{\ref@jnl{AcA}}
\newcommand\caa{\ref@jnl{ChA\&A}}
\newcommand\cjaa{\ref@jnl{ChJA\&A}}
\newcommand\jcap{\ref@jnl{JCAP}}
\newcommand\na{\ref@jnl{NewA}}
\newcommand\nar{\ref@jnl{NewAR}}
\newcommand\pasa{\ref@jnl{PASA}}
\newcommand\rmxaa{\ref@jnl{RMxAA}}

\title{High-frequency gravitational wave transients from superradiance}

\author[a]{Henry Su,}
\author[b,c]{Lucas Brown,}
\author[b,c]{Christopher Ewasiuk,}
\author[b,c]{and Stefano Profumo}

\affiliation[a]{Department of Physics, University of Massachusetts, Amherst, MA 01003, USA}
\affiliation[b]{Department of Physics, University of California, Santa Cruz \\Santa Cruz, CA, 95064, USA}
\affiliation[c]{Santa Cruz Institute for Particle Physics, University of California, Santa Cruz \\Santa Cruz, CA, 95064, USA}

\abstract{
Ultralight bosons can form macroscopic gravitational-atom clouds around rotating black holes via superradiance, sourcing quasi-monochromatic gravitational waves through level transitions and annihilation. Primordial black holes provide a natural setting for such systems in a frequency range relevant for resonant-cavity experiments. We present a unified treatment of gravitational-wave emission from both isolated and binary-perturbed gravitational atoms in this regime. For isolated systems, we derive analytic expressions for the time- and frequency-domain strain from transition and annihilation channels, emphasizing their narrow-band structure. For binaries, we model resonantly driven level transitions using the Landau--Zener formalism and compute the resulting narrow-band
signals. We find that binary-driven transitions generically yield signals with characteristic strains well below the sensitivity of current experiments at astrophysically plausible distances, and event rates further suppress detectability by requiring sources at unrealistically small separations. We quantify the improvements in sensitivity, bandwidth, and response needed to render these signals observable, and identify gravitational-atom systems around primordial black holes as a theoretically well-motivated target for future high-frequency gravitational-wave searches.
}

\begin{document}

\newcommand{\mBH}{m_{\text{BH}}}
\newcommand{\mBHseed}{\mBH^{\text{seed}}}
\newcommand{\mBHobs}{\mBH^{\text{obs}}}
\newcommand{\mBHobsi}{m_{\text{BH},i}^{\text{obs}}}
\newcommand{\mBHtheory}{\mBH^{\text{theory}}}
\newcommand{\zcoll}{z_{\text{coll}}}
\newcommand{\zvir}{z_{\text{vir}}}
\newcommand{\zobs}{z_{\text{obs}}}
\newcommand{\msun}{M_{\odot}}
\newcommand{\tsal}{t_{\text{sal}}}
\newcommand{\trel}{t_{\text{rel}}}
\newcommand{\rhocrit}{\rho_{\text{crit}}}
\newcommand{\cmg}{\text{cm}^{2}\text{g}^{-1}}
\newcommand{\kms}{\text{km}~\text{s}^{-1}}
\newcommand{\angstrom}{\r{A}}
\newcommand\sbullet[1][.5]{\mathbin{\vcenter{\hbox{\scalebox{#1}{$\bullet$}}}}}
\newcommand{\chisq}{\chi^{2}}
\newcommand{\Vmax}{V_{\text{max}}}
\newcommand{\lambdabar}{\bar{\lambda}}

\newcommand{\spr}[1]{{\color{red}\bf[SP:  {#1}]}}
\newcommand{\grantcomment}[1]{{\color{blue}\bf[GR:  {#1}]}}
\newcommand{\sg}[1]{{\color{red}\bf[SG:  {#1}]}}
\newcommand{\tesla}[1]{{\color{cyan}\bf[TJ:  {#1}]}}

\maketitle
\flushbottom

\section{Introduction}
\label{sec:intro}

The question of which physical mechanisms can produce gravitational waves (GWs) in
the MHz--GHz band has gained renewed urgency with the recent development of
resonant-cavity and broadband electromagnetic detectors sensitive to oscillatory
metric perturbations at these frequencies~\cite{BerlinHF,AggarwalHF,Domcke:2022}.
Unlike the audio-band signals detected by LIGO/Virgo/KAGRA or the mHz signals targeted
by LISA, high-frequency GWs cannot be sourced by stellar-mass compact binary mergers,
and the landscape of viable production mechanisms is far less explored.  Among the
handful of well-motivated theoretical candidates, black-hole superradiance stands out
because it can be made quantitative, is tied to motivated extensions of the Standard
Model, and---as we show in this paper---naturally places signals in the GHz band when
the black holes involved are of primordial origin and the boson masses are in a range motivated, for example, by the QCD axion phenomenology \cite{Peccei:1977hh, Weinberg:1977ma, Wilczek:1977pj,Preskill:1982cy, Abbott:1982af, Dine:1982ah,
diCortona:2015ldu}. We choose to use natural units
$\hbar = c = 1$ throughout this paper.

The basic mechanism is as follows.  A bosonic field of mass $\mu$ satisfying the
superradiant condition
\begin{equation}
  0 < \omega < m\,\Omega_H,
\end{equation}
where $\Omega_H$ is the angular velocity of the Kerr horizon and $m$ the azimuthal
quantum number, is exponentially amplified at the expense of the black hole's
rotational energy~\cite{Zeldovich:1971,Zeldovich:1972,PressTeukolsky:1972}.
When the boson's Compton wavelength is comparable to the gravitational radius, the
amplified modes are gravitationally trapped and grow into a macroscopic quasi-bound
cloud---a \textit{gravitational atom} (GA)~\cite{Detweiler:1980,Brito:2015}---with occupation numbers reaching
$N \sim 10^{70}$--$10^{80}$~\cite{Arvanitaki2015}.  The dynamics are
controlled by the gravitational fine-structure constant
\begin{equation}
  \alpha \equiv G M_{\rm BH}\,\mu,
\end{equation}
which simultaneously sets the cloud radius $r_c \sim M_{\rm BH}/\alpha^2$, the
superradiant instability rate, and the hydrogenic spectrum of bound
levels~\cite{ArvanitakiDubovsky,Arvanitaki2015}.  Efficient growth requires
$\alpha \sim 0.1$--$0.5$, which, for a given boson mass $\mu$, selects a preferred
BH mass---or equivalently, for a given BH mass, a preferred range of boson masses.

We emphasize that this relation should be interpreted as a target-region condition
rather than as a prediction that PBHs and an ultralight boson necessarily coexist
with the required masses.  For a given boson mass, efficient superradiance selects
the black-hole mass range for which $\alpha$ lies in the instability window;
conversely, for a given PBH mass, it selects the boson masses that can be probed.
The scenario considered below is therefore conditional on the existence of both
ingredients, and the resulting signals should be viewed as sharply defined targets
for high-frequency gravitational-wave searches rather than as generic predictions
of PBH cosmology.

The GW frequency is set almost entirely by $\mu$: annihilation of two cloud bosons
into a graviton emits at $\omega_{\rm ann} \simeq 2\mu$, while level transitions
between states with principal quantum numbers $n_g$ and $n_e$ emit at the Bohr
frequency $\omega_{\rm tr} \propto \mu\alpha^2(n_g^{-2}-n_e^{-2})$.  For stellar-mass
BHs this places signals in the audio band probed by LIGO/Virgo/KAGRA, and an
extensive literature has explored these GW signatures and the spin constraints they
imply~\cite{Brito:2015,Arvanitaki2015,Isi:2019,Ng:2019}.
A qualitatively distinct regime arises for \textit{primordial black holes}
(PBHs)~\cite{Carr:2020,Raidal:2017,AliHaemoud:2017}.  For sub-solar mass PBH masses the superradiance condition selects boson masses $\mu \sim 10^{-7}$--$10^{-5}$\,eV,
driving GW emission into the MHz--GHz band.  This is precisely the frequency range
targeted by resonant-cavity haloscopes such as ADMX, which were originally designed
to search for axion dark matter but can also be reinterpreted as high-frequency
gravitational-wave detectors~\cite{BerlinHF}.  For persistent, quasi-monochromatic
sources, the relevant comparison is to the steady-state cavity sensitivity, reached
when the signal duration exceeds the cavity response time.  For transient or
chirping sources, however, the response is instead determined by the convolution
of the source waveform with the cavity transfer function; the simple ring-up
condition used below should therefore be understood as a conservative diagnostic
rather than as a complete detector-response model~\cite{Schenk:2025cavity,Blas:2026multimode}.
 Also, notably, bosons in that mass range are independently motivated by the axion
solution to the strong CP problem~\cite{Peccei:1977hh, Weinberg:1977ma, Wilczek:1977pj}
and the cosmology thereof~\cite{Preskill:1982cy, Abbott:1982af, Dine:1982ah,
diCortona:2015ldu}.

Once a cloud has formed and superradiance saturates, two emission channels become
active in the isolated system:
\begin{enumerate}
  \item \textit{Level transitions}, in which the quadrupole self-interaction of
    two populated states drives bosons from an excited level to a lower one,
    emitting quasi-monochromatic GWs at the Bohr frequency $\omega_{\rm tr}$.
  \item \textit{Annihilations}, in which pairs of bosons convert into gravitons at
    $\omega_{\rm ann} \simeq 2\mu$, producing an extremely long-lived,
    slowly decaying quasi-monochromatic signal.
\end{enumerate}
These signals have been studied in detail for isolated stellar-mass GAs~\cite{Brito:2015,Arvanitaki2015,Isi:2019}, and we extend
that framework here to the PBH mass range and the associated GHz frequency
band.  Bosenova collapses, in which the
cloud implodes once its occupation number exceeds a critical value set by bosonic
self-interactions, introduce additional burst-like GW emission and periodically reset
the superradiant growth~\cite{YoshinoKodama:2012,YoshinoKodama:2014}; we account for
the resulting upper bound on cloud populations when deriving strain amplitudes.

In realistic astrophysical environments, however, GAs are often not isolated.  A
binary companion introduces a time-varying tidal potential that drives resonant
transitions between GA levels whenever the orbital frequency sweeps through a Bohr
frequency of the gravitational atom.  In this regime the cloud behaves as a driven
two-level system, and its response is captured by a Landau--Zener
analysis~\cite{Baumann2021}: depending on the adiabaticity parameter
$z \equiv \eta^2/(|\Delta m|\dot\Omega)$, where $\eta$ is the tidal mixing amplitude,
transitions range from partial population transfer ($z \ll 1$) to complete level
inversion ($z \gg 1$).  The resulting GW signal is a short-duration
burst whose characteristic strain and spectral shape have recently been computed by
Kyriazis \& Yang~\cite{Kyriazis:2025fis}, whose formalism we adopt and apply to
the PBH binary case.  Binary-induced resonant transitions also imply:
\begin{itemize}
  \item rapid depletion of the cloud, shortening or quenching the long-lived
    annihilation and transition signals;
  \item additional GW components at the resonance frequency,
    superimposed on the binary inspiral signal; and
  \item modifications to the binary's inspiral rate, including the possibility
    of ``floating orbits'' where cloud-mediated energy loss competes with GW
    radiation~\cite{Brito:2015,Baumann2021}.
\end{itemize}
Systematic studies of such binary-perturbed GAs have been carried out for stellar-mass
BH binaries and extreme mass-ratio inspirals relevant for
LISA~\cite{Baumann2020,Baumann2021,Tomaselli:2024}.  The present
work extends this program to PBH masses and GHz frequencies, a regime not
previously treated in a unified framework.

In this paper we bring together four ingredients that have not previously been
combined in the context of high-frequency GWs: (i)~relativistic superradiant
instability rates~\cite{Hoof_2025, Hoof2025Code}, (ii)~numerically computed cloud
masses~\cite{Siemonsen:2022yyf,May:2024}, (iii)~the binary-induced resonance
formalism of Ref.~\cite{Kyriazis:2025fis}, and (iv)~realistic ADMX sensitivity
curves~\cite{BerlinHF}.  Concretely, we make the following contributions:

\begin{itemize}
  \item We map the superradiant parameter space in the PBH mass range using
    Regge trajectories (Sec.~\ref{sec:Regge Trajectories}), identifying the
    $(\mu, M_{\rm BH}, \alpha)$ combinations that produce MHz--GHz emission and
    assessing the role of bosenova saturation and accretion in bounding the
    allowed cloud populations.

  \item For isolated GAs we derive time-domain GW strain envelopes and
    analytic closed-form frequency-domain templates, via the exponential
    integral $E_1$, for both the level-transition and annihilation signals
    (Sec.~\ref{sec:isolated-GA} and App.~\ref{sec:templates}).  We find that peak strains can reach $h \sim 10^{-22}$ at 1 kpc for the most
optimistic configurations of the annihilation channel, while level-transition
signals typically peak at $h \sim 10^{-23}$ and can be significantly smaller
for specific benchmark configurations.

  \item For PBH binaries we apply the Landau--Zener formalism to the hyperfine
    transition $\{2,1,1\} \to \{2,1,-1\}$, compute the short duration GW burst
    characteristic strain across the $M_{\rm BH}\text{--}\alpha$ parameter
    space (Sec.~\ref{sec:binary-GA}), and preform a
    conservative cavity-response diagnostic based on the relation between the signal duration
    and the cavity response time.

  \item We provide a detectability analysis for ADMX
    (Sec.~\ref{sec:binary-GA},~\ref{sec:pbh-rates}), showing that even under this conservative response diagnostic, the binary-induced burst
    strain falls orders of magnitude below the ADMX sensitivity threshold at
    astrophysically plausible distances.  Using PBH merger rates from the early
    two-body formation channel~\cite{Raidal:2024bmm}, we determine that
    detectable events would need to occur within hundreds of kilometers of the detector, whereas
    the merger rate implies characteristic event distances exceeding
    $\sim 1\,\text{kpc}$.
\end{itemize}

Taken together, our results show that PBH--gravitational atom systems are among the
very few theoretically well-motivated sources of MHz--GHz gravitational waves, but
that a significant improvement in detector strain sensitivity---beyond the reach of
current ADMX runs---is required to make them observable.  We identify the combination
of improved strain sensitivity, response models and readout strategies optimized
for transient and frequency-drifting sources, and broader low-frequency coverage
as the priority for future high-frequency GW detector design.  Our analytic frequency-domain
templates provide concrete waveform targets for such searches.

\section{Superradiance and the Parameter Space for High-Frequency GWs}
\label{sec:Regge Trajectories}

Rotating black holes interacting with ultralight bosonic fields define a
structured parameter space in which GW emission frequencies are tied
directly to the masses of both the black hole and the boson.  This section
establishes the parametric foundation for the rest of the paper.  We derive the
conditions for superradiant bound-state formation, introduce the Regge trajectories
that delimit the allowed region, give the instability rate governing cloud growth,
and identify the portion of parameter space that produces MHz--GHz gravitational
waves.  

The central result—summarized graphically in Fig.~\ref{fig:Regge}—is that the MHz–GHz window is
populated by black holes in the primordial mass range
$M_{\rm BH} \sim 10^{-6}-10^{-3}\,M_\odot$ for boson masses
$\mu \sim 10^{-8}-10^{-5}\,\mathrm{eV}$ and gravitational coupling $\alpha \sim 0.1-0.3$,
as appropriate for efficient superradiance.

\subsection{Bound States, Regge Trajectories, and the GW Frequency Map}
\label{sec:bound-states}

\begin{figure}[t]
\begin{center}
\includegraphics[width = 0.9\textwidth, height = 0.9\textwidth ]{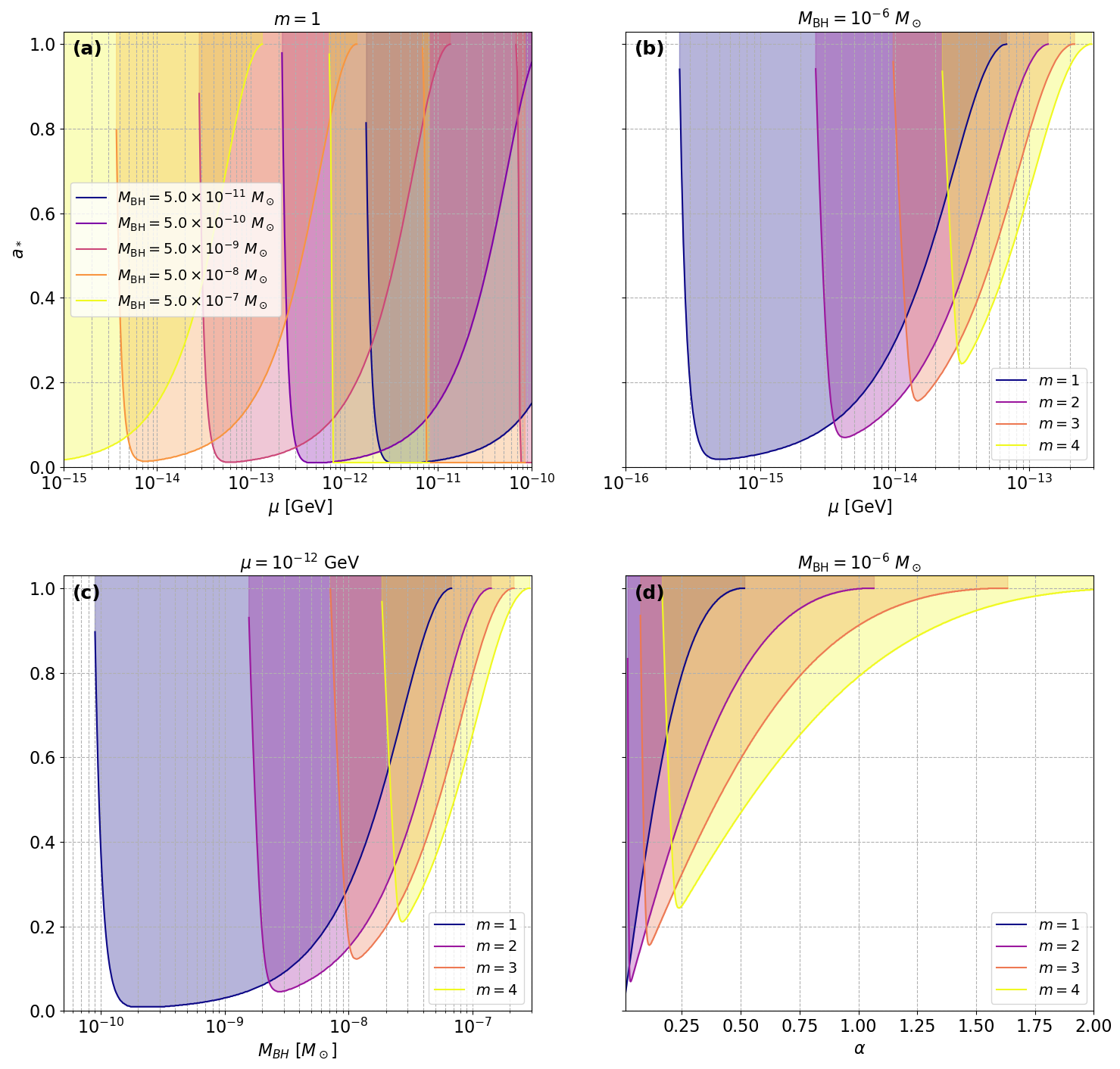}
\caption{Regge trajectories showing the minimum black hole spin $a_*$ required to sustain superradiant growth as a function of boson and black hole parameters, for azimuthal modes $m = 1, 2, 3, 4$. Shaded regions indicate parameter space where superradiance is active. \textbf{(a)} Superradiance threshold as a function of boson mass $\mu$ for fixed $m = 1$ and varying black hole mass $M$. \textbf{(b)} Threshold as a function of $\mu$ for fixed $M = 10^{-6}\ M_\odot$ and varying $m$. \textbf{(c)} Threshold as a function of $M_{\rm BH}$ for fixed boson mass $\mu = 10^{-3}\ \mathrm{eV}$ and varying $m$. \textbf{(d)} Threshold as a function of the gravitational coupling $\alpha = M\mu$ for fixed $M = 10^{-6}\ M_\odot$ and varying $m$.}
\label{fig:Regge}
\end{center}
\end{figure}

A massive scalar boson of mass $\mu$ on a Kerr background admits quasi-bound states
whenever its Compton wavelength $\lambdabar_C = \mu^{-1}$ (in natural units
$\hbar = c = 1$) is comparable to the gravitational radius $r_g \equiv GM_{\rm BH}$.
The ratio of these two scales is the gravitational fine-structure constant
\begin{equation}
    \alpha \;\equiv\; \frac{r_g}{\lambdabar_C} \;=\; G M_{\rm BH}\,\mu\,.
    \label{eq:alpha_def}
\end{equation}
In the limit $\alpha \ll 1$ the quasi-bound spectrum is precisely
hydrogenic~\cite{Detweiler:1980,Dolan:2007mj}:
\begin{equation}
    \omega_{n} \;\simeq\; \mu\!\left(1 - \frac{\alpha^2}{2n^2}\right),
    \label{eq:binding_energy}
\end{equation}
where $n$ is the principal quantum number.  The spatial extent of the cloud in
state $\{n, \ell, m\}$ is
\begin{equation}
    r_c \;=\; \frac{n^2}{\alpha^2}\,r_g \;=\; \frac{n^2}{\mu\alpha}\,,
    \label{eq:cloud_radius}
\end{equation}
so the cloud becomes more diffuse at smaller $\alpha$.  The regime $\alpha \sim
0.1$--$0.5$ simultaneously ensures that the cloud is compact enough for efficient
angular-momentum extraction and energetic enough for detectable GW emission; both
smaller and larger $\alpha$ values suppress the instability rate exponentially, as
discussed in Sec.~\ref{sec:SR-rates}.

Superradiance is active when the bound-state frequency satisfies the
\textit{superradiant condition}
\begin{equation}
    \omega < m\,\Omega_H\,,
    \label{eq:SR_condition}
\end{equation}
where $m$ is the azimuthal quantum number and
$\Omega_H = a_*/\!\left[2GM_{\rm BH}(1+\sqrt{1-a_*^2}\,)\right]$
is the angular velocity of the Kerr horizon~\cite{Brito:2015}, with $a_*$ the
dimensionless BH spin.  Substituting the leading-order binding energy
$\omega \approx \mu$ into Eq.~\eqref{eq:SR_condition} and solving for the
critical spin, one finds the threshold
\begin{equation}
    a_*^{\rm crit}(\alpha, m) \;=\; \frac{4m\alpha}{m^2 + 4\alpha^2}\,,
    \label{eq:a_crit}
\end{equation}
above which superradiance operates ($a_* > a_*^{\rm crit}$) and below which it
does not.  This threshold defines the \textit{Regge trajectory} of the mode: the
curve $a_* = a_*^{\rm crit}(\alpha, m)$ in the $(a_*, \alpha)$ plane, or
equivalently in the $(a_*, M_{\rm BH})$ or $(a_*, \mu)$ planes for fixed $\mu$
or $M_{\rm BH}$.  A BH that forms with spin above a Regge trajectory will be
driven toward it by superradiant spin-down; at the trajectory, growth saturates.

The spin requirement is an important model dependence of this construction.  In
the simplest radiation-dominated formation channel from nearly Gaussian curvature
perturbations, PBHs are generally expected to be born with small dimensionless spin,
because large overdensity peaks are close to spherical and pressure support
suppresses the growth of angular momentum~\cite{Chiba2017,DeLuca:2019pbhspin}.
However, larger spins can arise in less minimal formation histories.  Examples
include PBH formation during an early matter-dominated era, where tidal torques
and the longer collapse time can generate substantial angular momentum~\cite{Harada:2017pbhspin};
strongly nonspherical or non-Gaussian collapse; delayed collapse associated with
first-order phase transitions; and post-formation PBH--PBH mergers, whose remnants
naturally acquire spins of order $a_* \simeq 0.7$ for comparable-mass, initially
nonspinning binaries~\cite{Tichy:2008finalspin}.  Accretion can also spin up black
holes in sufficiently dense environments, although it is likely inefficient for
the isolated low-mass PBHs emphasized here.  We therefore do not assume that all
PBHs are rapidly rotating.  Rather, the Regge trajectories in Fig.~\ref{fig:Regge}
should be read as identifying the subset of PBHs whose spins exceed the relevant
threshold and hence can host the clouds studied in the rest of the paper.

In the limit $a_* \to 1$ (maximally spinning BH), Eq.~\eqref{eq:a_crit} gives
$a_*^{\rm crit} \to 1$ when $\alpha \to m/2$, so the hard upper bound for
superradiance to be possible at all is
\begin{equation}
    \alpha \;<\; \frac{m}{2}\,.
    \label{eq:alpha_upper}
\end{equation}
This bound is visible in all four panels of Fig.~\ref{fig:Regge} as the
right-hand edge of each shaded superradiant region.

Figure~\ref{fig:Regge} presents the Regge trajectories across four complementary
slices of parameter space.  In each panel the shaded area marks the region where a
given mode can grow:
\begin{itemize}
    \item \textit{Top-left}: the $(\mu, a_*)$ plane for varying $M_{\rm BH}$ at fixed
    $m=1$.  The allowed region shifts to larger boson masses as $M_{\rm BH}$
    decreases, illustrating that lighter PBHs couple to heavier bosons at the same
    $\alpha$.

    \item \textit{Top-right and bottom-left}: the $(\mu, a_*)$ and $(M_{\rm BH}, a_*)$ planes for varying $m$.  Higher azimuthal number extends the region to larger $\alpha$ (Eq. ~\eqref{eq:alpha_upper}); at fixed $\alpha$ these modes have a \textit{lower} critical spin but strongly suppressed growth rates, so $m=l=1$ dominates in practice

    \item \textit{Bottom-right}: the $(\alpha,a_*)$ plane at fixed $M_{\rm BH}$, showing the Regge trajectories in their most natural parametrization. The state with $m=\ell=1$ has the \textit{largest} critical spin of the four modes shown at given $\alpha$, but its superradiant growth rate is the least suppressed in $\alpha$ (Sec.~\ref{sec:SR-rates}); wherever the spin threshold is cleared, $m=\ell=1$ therefore grows fastest and extracts angular momentum first.
\end{itemize}
Throughout, quantum numbers are taken to satisfy $m = \ell = n - 1$ (the lowest
radial mode of each angular sector), since these states have the largest angular
momentum content per unit energy and therefore achieve the highest saturation
occupation numbers~\cite{Brito:2015,ArvanitakiDubovsky}.

The gravitational-wave emission frequencies are set by $\mu$ through two distinct
channels.  For \textit{annihilation} of two cloud bosons into a single graviton:
\begin{equation}
    \omega_{\rm ann}
    \;\simeq\; \frac{\mu}{\pi}
    \;\approx\; 484\,\mathrm{MHz}
    \left(\frac{\mu}{10^{-6}\,\mathrm{eV}}\right).
    \label{eq:f_ann}
\end{equation}
For a \textit{level transition} from principal quantum number $n_e$ down to $n_g$:
\begin{equation}
   \omega_{\rm tr}
    \;\simeq\; \frac{\mu\alpha^2}{4\pi}
    \!\left(\frac{1}{n_g^2} - \frac{1}{n_e^2}\right)
    \;\approx\; 4.84\,\mathrm{MHz}
    \left(\frac{\mu}{10^{-6}\,\mathrm{eV}}\right)
    \!\left(\frac{\alpha}{0.2}\right)^{\!2}
    \!\left(\frac{1}{n_g^2} - \frac{1}{n_e^2}\right).
    \label{eq:f_tr}
\end{equation}
The MHz--GHz frequency window therefore corresponds to boson masses
$\mu \sim 10^{-8}$--$10^{-5}$\,eV via annihilation, or to slightly heavier bosons
when transitions are the dominant channel.  The black hole mass required for
superradiance efficiency to peak at $\alpha \simeq 0.2$ is then
\begin{equation}
    M_{\rm BH} \;\simeq\; \frac{\alpha}{\mu\,G}
    \;\approx\; 2.7 \times 10^{-5}\,M_\odot
    \left(\frac{\alpha}{0.2}\right)
    \left(\frac{10^{-6}\,\mathrm{eV}}{\mu}\right),
    \label{eq:MBH_mu}
\end{equation}
Across the range $\mu \sim 10^{-8}-10^{-5}\,\mathrm{eV}$ relevant for MHz–GHz
emission, this corresponds to black hole masses
$M_{\rm BH} \sim 10^{-6}-10^{-3}\,M_\odot$ for $\alpha \sim 0.1-0.3$. This
is the PBH mass range, establishing the central connection exploited throughout
this paper: \textit{high-frequency GW detectors such as ADMX are natural instruments
for probing ultralight bosons through superradiance around PBHs.}  For reference,
Table~\ref{tab:freq_map} displays this mapping for frequencies spanning from low-RF band to the GHz range. This illustrates that the MHz–GHz frequency range associated with
superradiant gravitational atoms spans and extends beyond the current
ADMX sensitivity window, motivating broader frequency coverage in
future high-frequency gravitational-wave searches.

\begin{table}[t]
\centering
\caption{Mapping between boson mass $\mu$, host BH mass at $\alpha = 0.2$,
and annihilation GW frequency $\omega_{\rm ann}$.  The ADMX Run 1 data-collection band (0.645–1.4 GHz) lies between the
third and fourth rows, with the corresponding frequencies bracketing
the experimental sensitivity range.}
\label{tab:freq_map}
\renewcommand{\arraystretch}{1.35}
\begin{tabular}{|lccc|}
\hline
$\mu$ & $M_{\rm BH}/M_\odot$ & $\omega_{\rm ann}$ & Band \\
\hline
$10^{-9}$\,eV & $2.7\times10^{-2}$ & 484\, kHz       & low-RF     \\
$10^{-7}$\,eV & $2.7\times10^{-4}$ & 48\,MHz       & HF (broadband RF)    \\
$10^{-6}$\,eV & $2.7\times10^{-5}$ & 484\,MHz      & $\sim$ADMX Run~1 (lower)   \\
$3\times10^{-6}$\,eV & $9\times10^{-6}$ & 1.45\,GHz & ADMX Run~1 (upper)   \\
$10^{-5}$\,eV & $2.7\times10^{-6}$ & 4.8\,GHz      & Next-generation RF    \\
\hline
\end{tabular}
\end{table}

\subsection{Superradiant Instability Rates and Growth Timescales}
\label{sec:SR-rates}

The cloud occupation number grows as $\dot{N} = \Gamma_{n\ell m}\,N$, where
$\Gamma_{n\ell m} = \mathrm{Im}(\omega)$ is the superradiant instability rate.
In the non-relativistic limit ($\alpha \ll 1$) this rate can be computed
analytically by matched-asymptotic methods~\cite{Detweiler:1980,Brito:2015,Baumann_2019}:
\begin{equation}
\Gamma_{n\ell m} = \frac{2r_+}{M} C_{n\ell} g_{\ell m} (m\Omega_H - \omega_{n\ell m})\alpha^{4\ell + 5}, \label{eq:SR_rate} \\
\end{equation}
\text{with} 
\begin{equation*}
\quad C_{n\ell} = \frac{2^{4\ell + 1}(n + \ell)!}{n^{2\ell + 4}(n - \ell - 1)!} \left( \frac{\ell!}{(2\ell)! (2\ell + 1)!} \right)^2, \\
g_{\ell m} = \prod_{k = 1}^{\ell} \left[ k^2 (1 - {a^*}^2) + \left( a^* m - 2 r_+ \omega_{n\ell m} \right)^2 \right] 
\end{equation*}
where $C_{n\ell}g{_{\ell m}}(a_*)$ is a dimensionless, spin-dependent coefficient. Meanwhile, for a rotating Kerr black hole with angular momentum per unit mass \( a = J/Mc \), the event horizon is located at:

\begin{equation*}
r_+ = \frac{GM}{c^2} \left(1 + \sqrt{1 - \frac{a^2}{M^2 c^2}}\right).
\end{equation*}

\begin{figure}[t]
    \centering
    \begin{subfigure}{0.49\textwidth}
        \phantomcaption
        \label{fig:super_times_a}
    \end{subfigure}
    \begin{subfigure}{0.49\textwidth}
        \phantomcaption
        \label{fig:super_times_b}
    \end{subfigure}
    \begin{overpic}[width=\textwidth]{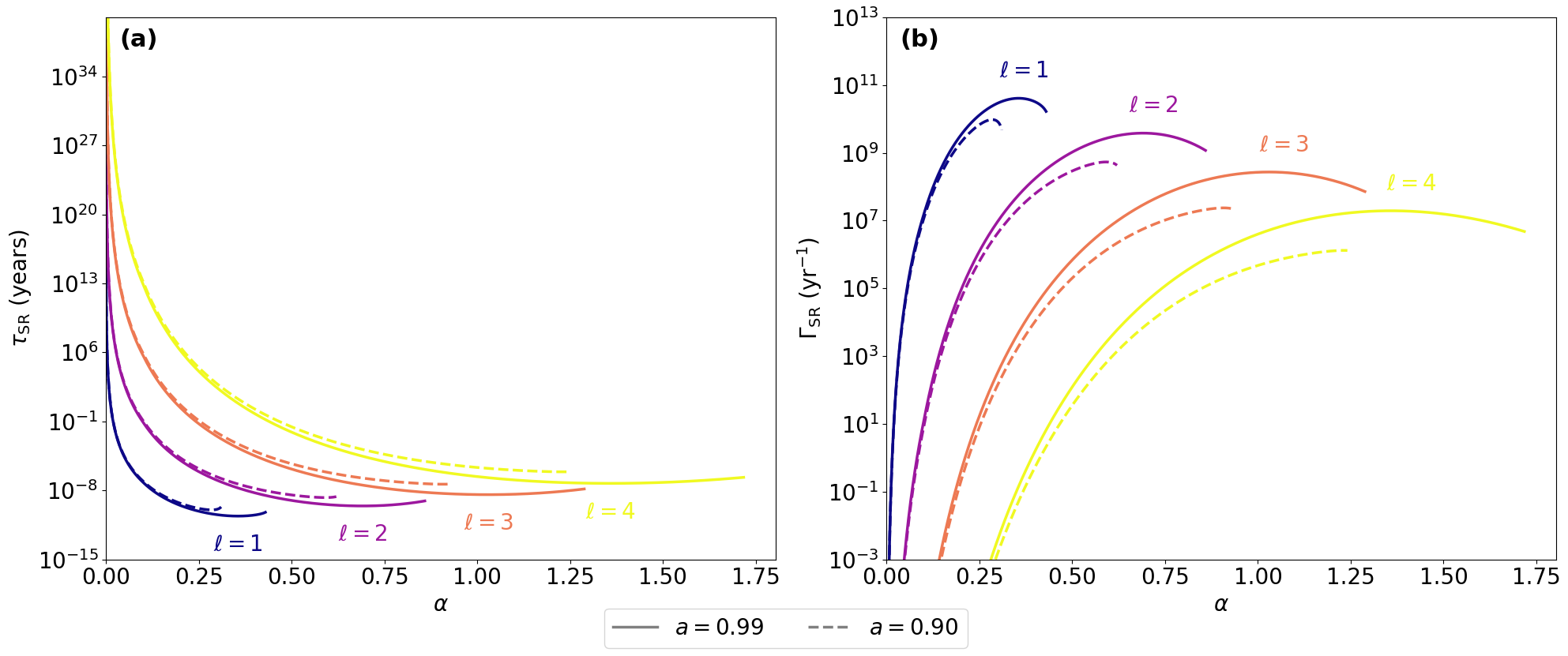}
    \end{overpic}
    \caption{\textbf{(a)} Estimates of superradiance lifetime $\tau = \Gamma^{-1}$ as a function of $\alpha$ and azimuthal quantum number $\ell = m$ for a $10^{-6}\,M_\odot$ BH. Solid lines show $a = 0.99$, dashed lines show $a = 0.90$. \textbf{(b)} Superradiance rate $\Gamma_{\rm SR}$ as a function of $\alpha$ for the same parameter space. Superradiance continues until $m\Omega_{\rm H} \geq \omega$ is violated through spin extraction via the superradiant cloud.}
    \label{fig:super_times}
\end{figure}

The \textit{e}-folding time for cloud growth is
\begin{equation}
    \tau_{\rm SR} \;\equiv\; \Gamma_{n\ell m}^{-1},
    \label{eq:tau_SR}
\end{equation}
Figure~\ref{fig:super_times_a} shows $\tau_{\rm SR}$ in years
as a function of $\alpha$ for modes $\ell = m=1,\ldots,4$; Fig.~\ref{fig:super_times_b}
shows the same rate in geometrized units $r_g\,\Gamma_{n\ell m}$, making the
$\ell$-dependent peak structure visible.

Three features of these figures are particularly relevant to our analysis:

\begin{enumerate}
\item
\textit{Steep $\alpha$-dependence at small $\alpha$.}\ Because
$\tau_{\rm SR} \propto \alpha^{-(4\ell+5)}$, the growth time rises by many orders
of magnitude as $\alpha$ decreases below $\sim 0.1$.  For the ADMX-relevant PBH
mass range, $\tau_{\rm SR}$ ranges from seconds at $\alpha \sim 0.4$ to timescales
exceeding the Hubble time at $\alpha \lesssim 0.05$.  This sets a practical lower
bound $\alpha \gtrsim 0.05$ for any detectable cloud to have formed.

\item
\textit{Cutoff at the superradiance threshold.}\ Both figures show $\Gamma_{n\ell m}
\to 0$ as $\alpha \to m/2$, because the superradiant enhancement $C_{n\ell}g_{\ell m}$
vanishes there.  The instability therefore operates within a finite window
$\alpha \in (0.05, m/2)$, peaking at $\alpha \sim 0.2$--$0.3$.

\item
\textit{Multi-mode competition.}\ For sufficiently high initial BH spin, several
$m$-modes can satisfy the superradiance condition simultaneously.  Although the
$m=1$ mode dominates the growth rate, modes with $m \geq 2$ develop non-negligible
populations over long timescales, contributing sub-dominant but non-zero GW
channels at their respective transition and annihilation frequencies.
\end{enumerate}

Throughout Sec. \ref{sec:isolated-GA} we compute instability rates from the matched-asymptotic expression of Eq.~\eqref{eq:SR_rate}, supplemented at larger $\alpha$ by the WKB envelope of Ref. ~\cite{ArvanitakiDubovsky} (their Eq.~27), since Eq.~\eqref{eq:SR_rate} alone is formally valid only for $\alpha \ll 1$.

\subsection{Cloud Saturation and Bosenova Cycling}
\label{sec:saturation}

Superradiant growth drives the BH spin from its initial value toward the Regge
trajectory $a_* = a_*^{\rm crit}$, at which point the instability saturates.
The fraction of BH mass transferred to the cloud at saturation is determined by
simultaneous conservation of energy and angular momentum.  For the dominant
$\{2,1,1\}$ state at $\alpha \simeq 0.2$ and $a_* \to 1$ this fraction is
approximately $10.8\%$~\cite{Tomaselli:2024}, consistent with our numerical
calculation of $10.81\%$ shown in Fig.~\ref{fig:Mass_fract}.  The corresponding
saturation occupation number is
\begin{equation}
N_{\rm sat} \approx \frac{G_N M_{\rm BH}^2\,\Delta a}{m} \approx 10^{65}\left(\frac{\Delta a}{0.1}\right)\left(\frac{M_{\rm BH}}{10^{-5}\,M_\odot}\right)^2.
    \label{eq:N_sat}
\end{equation}
The enormous value of $N_{\rm sat}$---a consequence of the tiny boson mass---is
what enables the macroscopic, coherent GW emission discussed in subsequent sections.

For bosons with attractive self-interactions, parametrized by the axion decay
constant $f_a$, a separate ceiling on the occupation number arises from the
\textit{bosenova} instability~\cite{ArvanitakiDubovsky,YoshinoKodama:2012}.  When
the occupation number exceeds
\begin{equation}
    N_{\rm bosenova} \;\approx\;
    5 \times 10^{78}
    \frac{n^4}{\alpha^3}
    \left(\frac{M}{M_\odot}\right)^{\!2}
    \left(\frac{f_a}{M_P}\right)^{\!2},
    \label{eq:N_bosenova}
\end{equation}
attractive self-interactions overcome the gravitational binding energy and the
cloud collapses~\cite{YoshinoKodama:2012,YoshinoKodama:2014}.  The implosion
deposits most of the extracted angular momentum back into the BH in a burst
lasting $\sim r_g/c$, after which superradiance restarts at reduced spin.  This
cycling behavior caps the effective occupation number at
$N_{\rm eff} \equiv \min(N_{\rm sat}, N_{\rm bosenova})$.

For QCD-axion values $f_a \sim 10^{16}\,\text{GeV}$ we find
$f_a/M_P \sim 10^{-3}$, giving
\begin{equation}
    \frac{N_{\rm bosenova}}{N_{\rm sat}} \;\sim\;
    5 \times 10^{6}
    \left(\frac{n^4/\alpha^3}{10}\right)
    \left(\frac{M_{\rm BH}}{10^{-5}\,M_\odot}\right)
    \left(\frac{\mu}{10^{-6}\,\mathrm{eV}}\right)
    \left(\frac{f_a/M_P}{10^{-3}}\right)^{\!2},
\end{equation}
so the bosenova threshold is not reached before saturation in the bulk of our
PBH parameter space.  We include $N_{\rm bosenova}$ as an upper bound in all
strain calculations, but note that for QCD-axion-like couplings it is
non-restrictive.  For very small $f_a$ (e.g., light moduli with $f_a \ll M_P$)
the bosenova may dominate and the cloud cycles; such scenarios, while physically
interesting, fall outside the scope of this paper.

Finally, we note that accreting matter onto the BH can spin it back toward
$a_* = 1$, potentially restarting superradiance after partial spin-down or
counteracting the spin-down while the cloud is growing.  For isolated PBHs in
vacuum—the scenario relevant to the PBH mass range considered here—negligible
accretion is expected and superradiance proceeds unimpeded.  The visible imprint
of accretion is the left-hand suppression of the allowed region in
Fig.~\ref{fig:Regge}: as $M_{\rm BH}$ decreases at fixed $\mu$, the condition
$\alpha < m/2$ becomes harder to satisfy if spin-up by accretion is competitive
with superradiant spin-down.  We treat this effect as absent in what follows.


\section{Gravitational-Wave Signatures from Isolated Gravitational Atoms}
\label{sec:isolated-GA}

Before examining the modifications introduced by a binary companion
(Sec.~\ref{sec:binary-GA}), we establish the GW signals from an \textit{isolated}
gravitational atom (GA) as the baseline against which binary-induced effects will
be compared.  Two physically distinct emission channels are active once superradiance
saturates:
\begin{enumerate}
  \item \textit{Level transitions} (Sec.~\ref{subsec:level_tr}), in which the
    gravitational self-interaction of two simultaneously occupied superradiant
    levels drives bosons from the higher to the lower state, emitting
    quasi-monochromatic GWs at the Bohr frequency $\omega_{\rm tr}$.
  \item \textit{Boson annihilation} (Sec.~\ref{subsec:annihilation}), in which pairs
    of cloud bosons annihilate into a single graviton at $\omega_{\rm ann} \simeq
    2\mu_a$, producing an extremely long-lived, slowly decaying signal.
\end{enumerate}
In both cases the signal is quasi-monochromatic and persists on timescales far
exceeding the ring-up time $\tau_{\rm ring}$ of a resonant-cavity detector; for these
persistent sources the steady-state cavity-sensitivity comparison is therefore the
appropriate limiting case.  The time-domain waveforms derived here
serve as inputs to the frequency-domain strain templates of App.~\ref{sec:templates}.

Throughout this section we assume the BH spin has been reduced to the saturation
value $a_*^{\rm crit}$ (Eq.~\eqref{eq:a_crit}) and that the cloud is in a
well-defined two-level configuration consisting of an ``excited'' state with quantum
numbers $(n_e, \ell, m)$ and a ``ground'' state $(n_g, \ell, m)$, with $n_e > n_g$.
All cloud masses are computed numerically using the results of
Refs.~\cite{Siemonsen:2022yyf, May:2024}, and superradiance rates use Eq ~\eqref{eq:SR_rate}.

\subsection{Gravitational Waves from Level Transitions}
\label{subsec:level_tr}

\begin{figure}[t!]
\begin{center}
\includegraphics[width=\textwidth,height=0.5\textwidth]{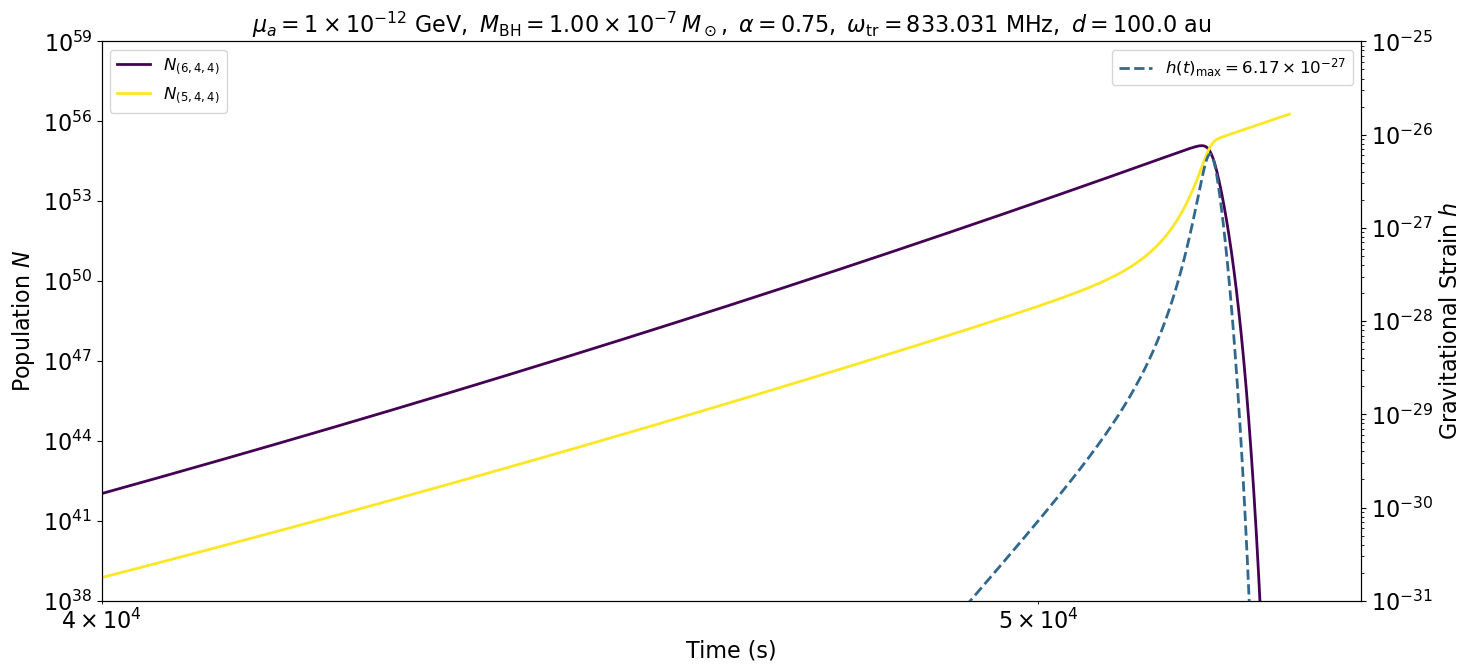}
\caption{Time evolution of the occupation numbers $N_{\{6,4,4\}}$  and
  $N_{\{5,4,4\}}$  alongside the corresponding gravitational-wave strain
  envelope $h(t)$ (dashed), computed for $\mu_a = 10^{-3}\,\text{eV}$,
  $M_{\rm BH} = 10^{-7}\,M_\odot$, $\alpha = 0.75$, and a source distance of
  $d = 100\,\text{au}$.  At early times both states grow independently under
  superradiance (Eqs.~\eqref{eq:dNg_indep}--\eqref{eq:dNe_indep}).  Once $N_e$
  is large enough, the gravitational self-interaction coupling $\Gamma_t N_g N_e$
  (Eq.~\eqref{eq:Gamma_t}) begins to dominate the excited-state dynamics, and a
  net flux of bosons cascades into the lower state, emitting a nearly monochromatic
  gravitational wave analogous to stimulated emission in atomic physics. The peak strain reflects a specific
benchmark configuration and is not representative of the maximal transition
strain, which can reach depends primarily on BH mass and source distance. This specific parameter configuration was chosen to illustrate a source within ADMX's detectable frequency band. }
\label{fig:level_strain}
\end{center}
\end{figure}

\begin{figure}[t!]
\begin{center}
\includegraphics[width=\textwidth,height=0.5\textwidth]{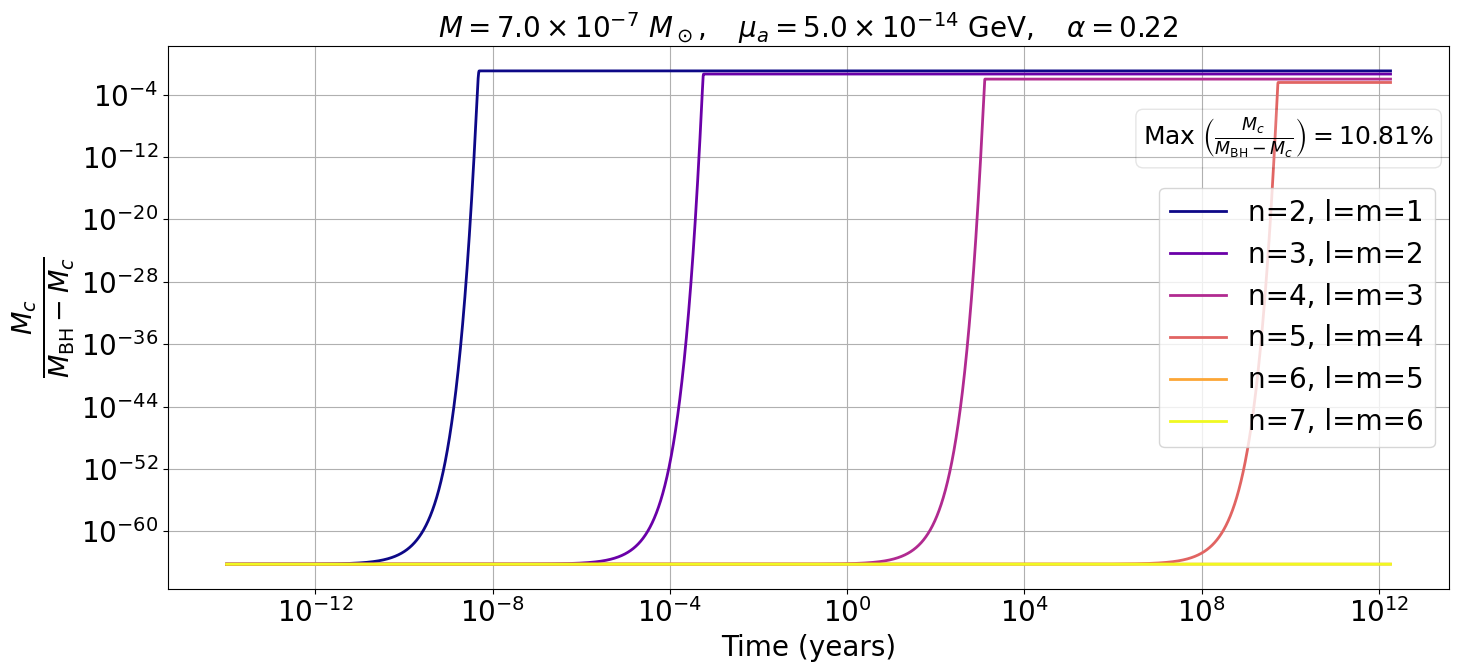}
\caption{Fractional BH mass transferred to each superradiant level,
  $M_c/(M_{\rm BH}-M_c)$, as a function of time for states with $n = m+1$,
  $\ell = m$, and $m = 1$--$6$, computed at $M_{\rm BH} = 6\times10^{-7}\,M_\odot$,
  $\mu_a = 5\times10^{-5}\,\text{eV}$, $\alpha = 0.22$.  The $\{2,1,1\}$ level
  (darkest) reaches its saturation first and at the largest fractional mass.  The
  maximum extracted mass for the most efficient state is $10.81\%$, in excellent
  agreement with the analytic upper bound of $10.8\%$ derived in
  Ref.~\cite{Tomaselli:2024} for $\alpha \approx 0.2$ and $a_* \to 1$.
 Higher-$m$ levels saturate later and accumulate correspondingly smaller clouds,
as their slower growth is overtaken by the already-saturated lower levels.}
\label{fig:Mass_fract}
\end{center}
\end{figure}

\subsubsection{Two-level population dynamics}
\label{subsubsec:population}

When the BH spin lies above the critical value for two levels simultaneously, both
states are initially populated independently through superradiance.  For
occupation numbers well below saturation, where the gravitational self-interaction
between levels is negligible, each level grows independently:
\begin{align}
  \frac{dN_g}{dt} &= N_g\,\Gamma_{\rm SR}^{(n_g)},
  \label{eq:dNg_indep}
  \\[4pt]
  \frac{dN_e}{dt} &= N_e\,\Gamma_{\rm SR}^{(n_e)},
  \label{eq:dNe_indep}
\end{align}
where $\Gamma_{\rm SR}^{(n)}$ is the superradiant instability rate for level $n$
(Eq.~\eqref{eq:SR_rate}).  The $\{6,4,4\}$ and $\{5,4,4\}$ states have nearly
equal instability rates at the benchmark parameters of Fig.~\ref{fig:level_strain},
so both grow on comparable timescales.

This picture changes once the occupation numbers become large enough that the
gravitational self-interaction of the cloud is no longer negligible.  A macroscopic
cloud in state $|n_e, \ell, m\rangle$ generates a time-dependent potential
$V_*(t, \bm{r})$ that mixes the two levels through the matrix element
$\langle n_g, \ell, m | V_*(t, \bm{r}) | n_e, \ell, m \rangle$.  This introduces a
coherent transition rate $\Gamma_t$, coupling the two occupation numbers:
\begin{align}
  \frac{dN_g}{dt} &= N_g\,\Gamma_{\rm SR}^{(n_g)} + N_g N_e\,\Gamma_t,
  \label{eq:dNg_coupled}
  \\[4pt]
  \frac{dN_e}{dt} &= N_e\,\Gamma_{\rm SR}^{(n_e)} - N_g N_e\,\Gamma_t.
  \label{eq:dNe_coupled}
\end{align}
The term $-N_g N_e\,\Gamma_t$ acts as a stimulated-emission drain on the excited
state: as $N_e$ grows, the transition rate accelerates, depopulating $|n_e\rangle$
and populating $|n_g\rangle$ while emitting a GW at the Bohr frequency.  The
per-boson transition rate is obtained from the gravitational quadrupole formula
applied to the off-diagonal current between the two
levels~\cite{ArvanitakiDubovsky,Arvanitaki2015}:
\begin{equation}
  \Gamma_t \;\sim\;
  \frac{2\,G_N\,\omega_{\rm tr}^5}{5}\,\mu_a^2\,r_c^4
  \;=\;
  \mathcal{O}(10^{-6}\text{--}10^{-8})\,\frac{G_N\,\alpha^9}{r_g^3}\,,
  \label{eq:Gamma_t}
\end{equation}
where $\omega_{\rm tr}$ is the transition frequency (Eq.~\eqref{eq:omega_tr}),
$r_c \sim n^2 r_g/\alpha^2$ is the characteristic cloud radius, and $r_g = GM$ is the
gravitational radius.  The scaling $\Gamma_t \propto \alpha^9/r_g^3$ follows from
substituting $r_c^4 \propto r_g^4/\alpha^8$ and $\omega_{\rm tr} \propto \mu_a\alpha^2
\propto \alpha^3/r_g$ into the first form.  The numerical range reflects the
variation across quantum numbers $n$ and $\ell$ for the configurations of
interest~\cite{ArvanitakiDubovsky}.

\subsubsection{Time-domain gravitational-wave strain}
\label{subsubsec:tr_strain}

The GW angular frequency for transitions between states $(n_e, \ell, m)$ and
$(n_g, \ell, m)$ with $n_e > n_g$ is the Bohr frequency:
\begin{equation}
  \omega_{\rm tr}
  \;=\;
  \omega_{n_g \ell m} - \omega_{n_e \ell m}
  \;\simeq\;
  \frac{\mu_a\,\alpha^2}{2}
  \left(\frac{1}{n_g^2} - \frac{1}{n_e^2}\right),
  \label{eq:omega_tr}
\end{equation}
where we have used the leading-order hydrogenic approximation.  For the $\{6,4,4\} \to \{5,4,4\}$ transition at the
benchmark parameters of Fig.~\ref{fig:level_strain}, this gives
$ \omega_{\rm tr}\approx 833.03\,\text{MHz}$.

The oscillating quadrupole moment of the two-level system sources a GW whose strain envelope follows from the standard power–strain relation $h^2 = 4G_N P_{\rm tr}/(r^2\omega_{\rm tr}^2)$ [16]. Since each transition quantum carries energy $\omega_{\rm tr}$ and the quanta are emitted at rate $\Gamma_t N_g N_e$, the radiated power is $P_{\rm tr}=\omega_{\rm tr}\,\Gamma_t N_g(t) N_e(t)$; substituting and canceling one power of $\omega_{\rm tr}$ gives: ~\cite{ArvanitakiDubovsky,Arvanitaki2015}:
\begin{equation}
  h_{0,\rm tr}(t)
  \;=\;
  \sqrt{\frac{4\,G_N}{r^2\,\omega_{\rm tr}}\,\Gamma_t\,N_g(t)\,N_e(t)}\,,
  \label{eq:transition_envelope}
\end{equation}
where $N_g(t)$ and $N_e(t)$ are solutions of
Eqs.~\eqref{eq:dNg_coupled}--\eqref{eq:dNe_coupled}.  This is the
\textit{slowly-varying envelope} of the GW signal: it evolves on timescales set by
$\Gamma_{\rm SR}$ and $\Gamma_t$, which are much longer than the GW oscillation period
$2\pi/\omega_{\rm tr}$.  The full waveform, decomposed into the standard
plus and cross polarizations, is:
\begin{align}
  h_+(t) &= h_{0,\rm tr}(t)\,
    \frac{1+\cos^2\!\iota}{2}\,
    \cos\!\big(\omega_{\rm tr}\,t + \phi_0\big),
  \label{eq:h_tr_plus}
  \\[4pt]
  h_\times(t) &= h_{0,\rm tr}(t)\,
    \cos\iota\,
    \sin\!\big(\omega_{\rm tr}\,t + \phi_0\big),
  \label{eq:h_tr_cross}
\end{align}
where $\iota$ is the inclination of the BH spin axis with respect to the line of
sight and $\phi_0$ is an arbitrary initial phase.  The signal is quasi-monochromatic
at $\omega_{\rm tr}$ with a slowly drifting amplitude; it is the amplitude modulation
through $h_{0,\rm tr}(t)$ that, in the frequency domain (App.~\ref{sec:templates}),
produces the characteristic narrow Lorentzian lineshape at $f = \omega_{\rm tr}/2\pi$.

\subsubsection{Saturation occupation number and peak strain}
\label{subsubsec:tr_peak}

The peak GW strain occurs at the moment when the coupled system
(Eqs.~\eqref{eq:dNg_coupled}--\eqref{eq:dNe_coupled}) transitions from the
superradiance-dominated to the transition-dominated regime, i.e., when
$\Gamma_t N_e \sim \Gamma_{\rm SR}^{(n_e)}$. This equal population condition corresponds to the moment of maximum stimulated emission and thus maximum GW power. Specifically, the maximum gravitational wave power occurs when the excited state equals the ground state
\begin{equation}
    N_{tr} = \frac{\Gamma_{n\ell m} }{\Gamma_{t}}.
\end{equation}

Note that this level transition always occurs before the cloud population becomes saturated. In Fig. \ref{fig:level_strain}, the transition occurs when $N_g = N_e = 10^{55}$, and the population continues to grow until $N_{sat} \approx 10^{62}$. Substituting $N_g \sim N_e \sim N_{\rm tr}/2$
into Eq.~\eqref{eq:transition_envelope} gives the peak strain:
\begin{equation}
  h_{0,\rm tr}^{\rm peak}
  \;\simeq\;
  \frac{1}{2}\,
  \sqrt{\frac{4\,G_N\,\Gamma_{n\ell m}^2 \,}{r^2\,\omega_{\rm tr}\Gamma_t}}
  \;\simeq\;
  8\times10^{-31}
  \left(\frac{1\,\text{au}}{r}\right)
  \left(\frac{\alpha}{0.3}\right)^{15}
  \left(\frac{M}{10^{-7}\,M_\odot}\right).
\end{equation}
where the numerical scaling absorbs $G_N$, $r_g$, and $N_{\rm sat}(\alpha, M)$, and is
normalized to the $\ell = 4$ transition ($\{6,4,4\}\to\{5,4,4\}$) shown in
Fig.~\ref{fig:level_strain}. Higher-$\ell$ transitions are far weaker than the
dominant $\ell = 1$ channel: the wavefunction overlap entering $\Gamma_{n\ell m}$
and $\Gamma_t$ suppresses $h^{\rm peak}_{0,\rm tr}$ by roughly two additional orders
of magnitude per unit of $\ell$, and steepens the $\alpha$-scaling from $\alpha^3$
at $\ell = 1$ to $\alpha^{15}$ here. This scaling should be treated as an
order-of-magnitude guide rather than a precise formula, since the coefficient
also depends sensitively on the black-hole spin near the superradiant threshold.

Using the benchmark parameters of $M = 10^{-7}\,M_\odot$ and $\alpha = 0.75$ we
obtain $h^{\rm peak}_{0,\rm tr}\approx 7\times10^{-27}$ at $r=100$ au, in good
agreement with the numerically computed peak $h_{\max} \approx 6\times10^{-27}$
seen in Fig.~\ref{fig:level_strain}. The smallness of this value relative to the
optimal $\ell=1$ case reflects the $\ell = 4$ level structure adopted there,
rather than the maximal achievable transition strain.

\subsubsection{Signal lifetime}
\label{subsubsec:tr_lifetime}

Once the transition-dominated phase begins, the excited state drains with a
characteristic timescale
\begin{equation}
  \tau_{\rm tr}
  \;\sim\;
  \frac{1}{\Gamma_t\,N_{\rm sat}}
  \;\sim\;
  \mathcal{O}(10^3\text{--}10^6)\;\text{yr},
  \label{eq:tau_tr}
\end{equation}
for the parameters relevant to MHz--GHz emission.  Since $\tau_{\rm tr} \gg
\tau_{\rm ring} \sim 1/\Delta f_{\rm band} \sim 10^{-5}\,\text{s}$ (for a cavity
bandwidth $\Delta f_{\rm band} \sim 10\,\text{kHz}$), the level-transition signal
is effectively persistent on detector timescales.  The extremely narrow
spectral width $\Delta f \sim \tau_{\rm tr}^{-1}/(2\pi) \ll 1\,\text{Hz}$
(compared to the carrier frequency $f_{\rm tr} \sim \text{MHz--GHz}$) makes this
signal highly coherent over any realistic observation window; the resulting
frequency-domain Lorentzian template is developed in App.~\ref{sec:templates}.

\subsection{Gravitational Waves from Boson Annihilation}
\label{subsec:annihilation}

\begin{figure}[t!]
\begin{center}
\includegraphics[width=0.9\textwidth,height=0.5\textwidth]{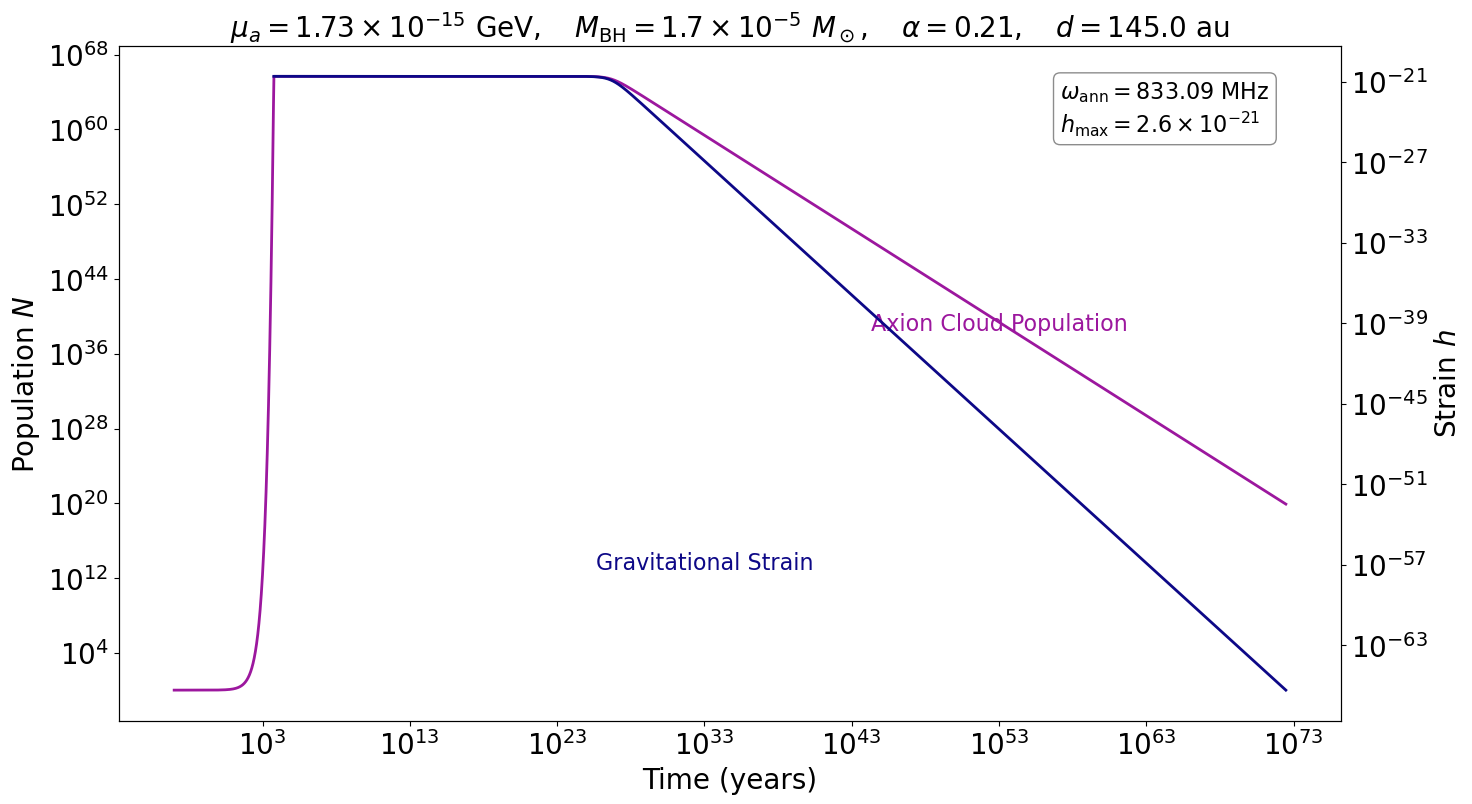}
\caption{Time evolution of the cloud occupation number $N(t)$ (solid) and the
  corresponding annihilation strain envelope $h_{0,\rm ann}(t)$ (dashed), computed for
  $\mu_a \approx 2\times10^{-6}\,\text{eV}$, $M_{\rm BH} = 1.7\times10^{-5}\,M_\odot$,
  $\alpha = 0.21$, and $d = 145\,\text{au}$.  The occupation number grows during the
  superradiant phase (left portion of the curve) until it saturates at
  $N_{\rm max}$, after which $N(t)$ decays as $[1 + \Gamma_a N_{\rm max}\,t]^{-1}$
  (Eq.~\eqref{eq:N_ann}) as boson pairs annihilate into gravitons.  The
  corresponding strain peaks at $h_{0,\rm ann}^{\rm max} \approx 10^{-21}$ for nearby distances and
  decays on a timescale $\tau_{\rm ann} = (\Gamma_a N_{\rm max})^{-1}$ that can
  exceed the Hubble time, making the signal effectively monochromatic and
  persistent for any observation campaign.  The annihilation frequency
  $\omega_{\rm ann}/(2\pi) \approx 833.09 \text{ MHz}$ was chosen to match the emitted signal from Fig. \ref{fig:level_strain} and is within the ADMX scan band. }
\label{fig:annihilation_strain}
\end{center}
\end{figure}

\subsubsection{Annihilation frequency and rate}
\label{subsubsec:ann_rate}

When two cloud bosons annihilate they produce a single graviton.  Each boson in
the bound state has energy $\omega_b = \omega_{n\ell m}$, so the emitted graviton carries
\begin{equation}
  \omega_{\rm ann}
  \;=\;
  2\,\omega_{n\ell m}
  \;\simeq\;
  2\mu_a\!\left(1 - \frac{\alpha^2}{2n^2}\right),
  \label{eq:omega_ann}
\end{equation}
to leading order in $\alpha$.  The $\mathcal{O}(\alpha^2)$ correction shifts the
emission frequency slightly below $2\mu_a$; for the benchmark parameters of
Fig.~\ref{fig:annihilation_strain} this gives $f_{\rm ann} = \omega_{\rm ann}/(2\pi)
\approx 833.09\,\text{MHz}$.

The pair-annihilation rate per boson—the rate at which a single pair converts
into a graviton—is obtained from the gravitational quadrupole formula applied to
the $2\mu_a$ oscillation of the cloud's stress--energy
tensor~\cite{Arvanitaki2015}:
\begin{equation}
  \Gamma_a
  \;\simeq\;
  K_{n\ell m}
  \left(\frac{\alpha/\ell}{0.5}\right)^{p}
  \frac{G_N}{r_g^3}\,,
  \label{eq:Gamma_a}
\end{equation}
where the exponent $ p = 17$ for $\ell = 1$ and $p = 4\ell + 11$ for $\ell \geq  2$  is determined by the power of $\alpha$ in
the hydrogenic wavefunction overlap, and $K_{n\ell m} \approx 10^{-10}$ is a
dimensionless coefficient that depends weakly on $n$ for $n \lesssim \ell +
2$~\cite{Arvanitaki2015,Brito:2015}.  Since
$\Gamma_a \propto \alpha^{4\ell+11}/r_g^3$ while $\Gamma_{\rm SR} \propto
\alpha^{4\ell+5}/r_g$ and $\Gamma_t \propto \alpha^9/r_g^3$, the hierarchy
$\Gamma_{\rm SR} \gg \Gamma_t \gg \Gamma_a$ holds over the entire parameter space of
interest, confirming that annihilation is the slowest of the three processes.

\subsubsection{Time-domain occupation number and strain}
\label{subsubsec:ann_strain}

After superradiance saturates the cloud at $N_{\rm max}$ and the level-transition
phase has depleted the excited state, the occupation number of the surviving ground
level evolves under pure annihilation.  Solving $dN/dt = -\Gamma_a N^2$ gives:
\begin{equation}
  N(t)
  \;=\;
  \frac{N_{\rm max}}{1 + \Gamma_a\,N_{\rm max}\,t}\,,
  \label{eq:N_ann}
\end{equation}
starting from $t = 0$ at saturation.  The GW strain envelope follows directly from
the quadrupole luminosity:
\begin{equation}
  h_{0,\rm ann}(t)
  \;=\;
  N(t)\,
  \sqrt{\frac{4\,G_N}{r^2\,\omega_{\rm ann}}\,\Gamma_a}
  \;=\;
  \frac{N_{\rm max}}{1 + \Gamma_a N_{\rm max} t}\,
  \sqrt{\frac{4\,G_N\,\Gamma_a}{r^2\,\omega_{\rm ann}}}\,,
  \label{eq:h0ann_env}
\end{equation}
which is the slowly-varying amplitude of the wave~\cite{Arvanitaki2015}.  The full
waveform polarizations are:
\begin{align}
  h_+(t)
  &= h_{0,\rm ann}(t)\,
    \frac{1+\cos^2\!\iota}{2}\,
    \cos\!\big(\omega_{\rm ann}\,t + \phi_0\big),
  \label{eq:h_ann_plus}
  \\[4pt]
  h_\times(t)
  &= h_{0,\rm ann}(t)\,
    \cos\iota\,
    \sin\!\big(\omega_{\rm ann}\,t + \phi_0\big),
  \label{eq:h_ann_cross}
\end{align}
where the inclination $\iota$ and phase $\phi_0$ carry the same meaning as in
Eqs.~\eqref{eq:h_tr_plus}--\eqref{eq:h_tr_cross}.  The envelope $h_{0,\rm ann}(t)$
includes the effects of BH mass and spin loss, which cause $N_{\rm max}$ and
$\omega_{\rm ann}$ to decrease slowly over time; these corrections are
numerically small ($\lesssim 10\%$ over cosmic timescales for the configurations
considered) and are incorporated in our numerical calculations.

\subsubsection{Peak strain scaling}
\label{subsubsec:ann_peak}

The peak strain, attained at $t = 0$ (saturation), is:
\begin{equation}\label{eq:hpeak_ann}
\begin{split}
  h_{0,\rm ann}^{\rm peak}
  \;=\;
  N_{\rm max}\,
  \sqrt{\frac{4\,G_N\,\Gamma_a}{r^2\,\omega_{\rm ann}}}
  &\;\simeq\; 
  2 \times10^{-20}
  \left(\frac{1\,\text{au}}{r}\right)
  \left(\frac{\alpha/\ell}{0.5}\right)^{p/2}
  \left(\frac{M}{10^{-6}\,M_\odot}\right) \\
  &p = 17\ (\ell=1),\ \ 4\ell+11\ (\ell\geq2),
\end{split}
\end{equation}

where the numerical scaling uses $N_{\rm max} \sim M^2(\tilde{a}_*^{\rm init}-a_*^{\rm crit})
\cdot \omega_{n\ell m}/m$ and the $\Gamma_a$ scaling of Eq.~\eqref{eq:Gamma_a}. For the benchmark parameters of
Fig.~\ref{fig:annihilation_strain}, the simulated peak is $h_{0,\rm ann}^{\rm max}
\approx 10^{-21}$, consistent with this estimate.

\subsubsection{Signal lifetime and detectability window}
\label{subsubsec:ann_lifetime}

The characteristic decay time of the annihilation signal is:
\begin{equation}
\tau_{\rm ann} \equiv \frac{1}{\Gamma_a N_{\rm max}} \sim \frac{r_g^3}{G_N K_{n\ell m}}\left(\frac{0.5}{\alpha/\ell}\right)^{p}\frac{1}{N_{\rm max}},
  \label{eq:tau_ann}
\end{equation}
which, for the parameters relevant to MHz--GHz emission, evaluates to
$\tau_{\rm ann} \sim 10^{25}$--$10^{60}\,\text{yr}$.  This vastly exceeds both the
Hubble time and any conceivable observation period, confirming that the annihilation
signal is effectively a \textit{monochromatic steady source} for the purposes of
detection.  The extremely slow decay also means that $N(t) \approx N_{\rm max}$
throughout any observation, so $h_{0,\rm ann}$ is well approximated by its peak value.

Like the transition signal, the annihilation signal is continuous on detector
timescales ($\tau_{\rm ann} \gg \tau_{\rm ring}$), and its spectral width
$\Delta f \sim \tau_{\rm ann}^{-1}/(2\pi) \lesssim 10^{-33}\,\text{Hz}$ is
extraordinarily narrow.  The frequency-domain template, detailed in
App.~\ref{sec:templates}, takes the form of an exponential-integral
function peaked at $f_{\rm ann} = \omega_{\rm ann}/(2\pi)$ with a width set by
$\Gamma_a N_{\rm max}$.

\subsection{Summary and comparison of isolated-GA signals}
\label{subsec:isolated_summary}

Table~\ref{tab:isolated_summary} collects the key properties of the two isolated-GA
GW channels for the benchmark parameter sets introduced in Table~\ref{tab:freq_map}.
The level-transition and annihilation signals differ in amplitude, duration, and
spectral width, but share the property that both persist for times enormously
exceeding the ring-up time of any resonant-cavity detector.  This motivates the
use of characteristic strain $h_c(f) = 2f|\tilde{h}(f)|$ as the figure of merit
(App.~\ref{sec:templates}), since the large number of coherent cycles
accumulated over $\tau_{\rm tr}$ or $\tau_{\rm ann}$ enhances the characteristic
strain above the raw amplitude by a factor $\sqrt{f\,\tau}$.

\begin{table}[t!]
\centering
\renewcommand{\arraystretch}{1.4}
\begin{tabular}{|lcc|}
\hline
Property & Level transition & Annihilation \\
\hline
Emission frequency & $\omega_{\rm tr}/(2\pi)$, Eq.~\eqref{eq:omega_tr}
                   & $\omega_{\rm ann}/(2\pi)$, Eq.~\eqref{eq:omega_ann} \\
Rate controlling amplitude & $\Gamma_t$, Eq.~\eqref{eq:Gamma_t}
                           & $\Gamma_a$, Eq.~\eqref{eq:Gamma_a} \\
Peak strain at 1\,kpc & $\sim 10^{-23}$ & $\sim 10^{-22}$ \\
Signal lifetime & $\tau_{\rm tr}\sim10^3$--$10^6\,\text{yr}$,
                  Eq.~\eqref{eq:tau_tr}
                & $\tau_{\rm ann}\sim10^{25}$--$10^{60}\,\text{yr}$,
                  Eq.~\eqref{eq:tau_ann} \\
Spectral width $\Delta f$ & $\sim \tau_{\rm tr}^{-1}/(2\pi) \ll 1\,\text{Hz}$
                          & $\lesssim 10^{-33}\,\text{Hz}$ \\
$\Delta f/f$ & $\ll 10^{-6}$ & $\ll 10^{-30}$ \\
Signal type & Quasi-monochromatic transient & Persistent monochromatic \\
\hline
\end{tabular}
\caption{Comparison of the two isolated-GA gravitational-wave channels for the
  PBH benchmark parameters of Table~\ref{tab:freq_map}.  Both signals satisfy
  $\tau \gg \tau_{\rm ring}$ and are candidates for resonant-cavity
  detection in the MHz--GHz band.  The level-transition signal is stronger at
  early times but decays over $\tau_{\rm tr}$; the annihilation signal is weaker
  but essentially eternal.}
\label{tab:isolated_summary}
\end{table}

The key conclusion of this section is that \textit{both} isolated-GA emission channels
produce continuous, narrowband GWs in the MHz--GHz band for PBH masses in the range
$M_{\rm BH} \sim 10^{-16}$--$10^{-4}\,M_\odot$.  Their detectability at a specific
detector depends on the signal duration relative to $\tau_{\rm ring}$ (which is
easily satisfied) and on whether the peak strain exceeds the detector noise floor.

\section{Gravitational-Wave Signatures from Gravitational Atoms in Binaries}
\label{sec:binary-GA}

The emission channels studied in Sec.~\ref{sec:isolated-GA} treat the gravitational
atom (GA) as evolving in isolation.  In realistic astrophysical environments, however,
a GA may reside in a binary system whose companion's tidal field drives resonant
transitions between cloud levels.  When the orbital frequency $\Omega_0$ sweeps through
a Bohr frequency of the gravitational atom, the system behaves as a driven two-level
quantum system and can undergo a Landau--Zener resonance~\cite{Baumann2020,Baumann2021}.
In this section we apply the binary-perturbation formalism of Kyriazis \& Yang~\cite{Kyriazis:2025fis}
to PBH binaries in natural units $c = \hbar = 1$, compute the characteristic GW strain
of the transient burst produced by the resonant $\{2,1,1\} \to \{2,1,-1\}$ hyperfine
transition, and assess the prospects for detection with ADMX.

\subsection{Time- and Frequency-Domain Waveforms}
\label{subsec:gab_waveform}

The gravitational-wave strain produced by tidal perturbations of a GA cloud from
its binary companion can be expressed as~\cite{Kyriazis:2025fis}
\begin{align}
  h_{+,211}(t) &= h_0\,
    \frac{1+\cos^2\iota}{2}\,
    \Re\!\left[e^{-2i\Delta m\,\varphi(t_{\rm re})}\,Q(t_{\rm re})\right],
  \label{eq:binary_signal_h+}
  \\[4pt]
  h_{\times,211}(t) &= h_0\,
    \cos\iota\,
    \Im\!\left[e^{-2i\Delta m\,\varphi(t_{\rm re})}\,Q(t_{\rm re})\right],
  \label{eq:binary_signal_hx}
\end{align}
where $\iota$ is the inclination angle between the line of sight and the normal to
the binary orbital plane, and $t_{\rm re}$ is the retarded time.  The oscillating term $e^{-2i\Delta m\varphi(t_{\rm re})}$ represents the carrier wave, where $-2\Delta m\varphi(t_{\rm re}) = \Phi_{GW}(t)$ represents the corresponding GW phase. The binary orbital
phase is $\varphi(t) = \Omega_0 t + \tfrac{1}{2}\gamma t^2 + \phi_0$, which assumes a
linearly chirping orbit with rate $\dot\Omega = \gamma$ (see Eq.~\eqref{eq:gamma_chirp}
below) within an arbitary phase factor of $\phi_0$. The instantaneous GW carrier frequency is then defined by $\dot \Phi_{GW}(t)$ = $-2\Delta m\Omega_{0} -2 \Delta m\gamma t$, resembling a linear chirping signal. The quantity $\Delta m \equiv m_f - m_i = -2$ is the difference in azimuthal
quantum numbers between the final ($m_f = -1$) and initial ($m_i = +1$) states of
the $\{2,1,1\}\to\{2,1,-1\}$ transition. The complex modulation $Q(t)$ captures the
dynamics of the level population transfer: it encodes the evolving occupation-number
amplitudes of the two mixed states as the binary sweeps through resonance, and
its derivation and explicit form are given in Ref.~\cite{Kyriazis:2025fis}.  The
exponential factor $e^{-2i\Delta m\,\varphi}$ carries the fast orbital-phase
oscillation, while the variation of $Q(t)$ is slow compared to $1/\Omega_0$. 

The overall GW amplitude is set by
\begin{equation}
  h_0 = \frac{24\,q_c\,GM}{r}\,\frac{(GM\Omega_0)^2}{\alpha^4}\,,
  \label{eq:h0_factor}
\end{equation}
where $M$ is the mass of the host BH (the primary of the gravitational-atom--binary,
or GAB, system), $\Omega_0$ is the binary orbital frequency at resonance, $r$ is the
observer distance, and $q_c \equiv \frac{M_{\rm cloud}}{M-M_{\rm cloud}}$ is the cloud-to-BH mass ratio.
The latter is determined by the saturation condition~\cite{Kyriazis:2025fis}:
\begin{equation}
  q_c = \frac{8\alpha^2\!\left(1 - \dfrac{\alpha a_{*}}{m}\right)}
    {m^2\!\left(1 - \sqrt{1 - \left(\dfrac{4\alpha}{m}
      \left(1 - \dfrac{\alpha a_{*}}{m}\right)\right)^{\!2}}\,\right)} - 1\,,
  \label{eq:qc}
\end{equation}
where $m$ is the azimuthal quantum number of the initial state and
$a_*$ is the dimensionless BH spin.  Throughout we set $a_* =
a_{*}^{\rm crit}(m_i)$ (Eq.~\eqref{eq:a_crit}), i.e., the spin is taken
at the Regge-trajectory saturation value that maximizes $\alpha$ for fixed
boson mass.

For the rest of the section, we shall only concern ourselves with the “plus” polarization $h_{+}$, as the two strain polarizations only differ by a relative phase and inclination weight, which can be observed through the comparison between Eq. {\eqref{eq:binary_signal_h+}}  and Eq. {\eqref{eq:binary_signal_hx}} for the time domain, and proved within Section {\ref{sec:binary template}} for the frequency domain. 

In frequency space, the Fourier transform of $h_+(t)$ takes the analytic
form~\cite{Kyriazis:2025fis}
\begin{equation}
  \tilde{h}_+(f) = h_0(1+\cos^2\iota)\,\sqrt{\pi}\,|\Delta m|^2\,i\,
    e^{i\Psi_+(f)}\,
    \frac{\sqrt{z}}{|\Gamma| - i\pi(f-f_c)}\,
    e^{-\pi z}\,
    \exp\!\left[-2z\arctan\!\left(\frac{\pi(f-f_c)}{|\Gamma|}\right)\right]\,
  \label{eq:binary_h_plus_fourier}
\end{equation}
where $f_c = 2\Omega_0/\pi$ is the carrier frequency (four times the orbital frequency
divided by $2\pi$), and the GW phase is
$\Psi_+(f) = 2\pi fr + 4\pi^2(f-f_c)^2/(4|\Delta m|\gamma) - \pi/4$.
The quantity $\Gamma \equiv \Gamma_{\rm SR}^{(2,1,-1)}$ is the superradiant
instability rate of the \textit{final} state $\{2,1,-1\}$ (Eq.~\eqref{eq:SR_rate}),
which governs how rapidly bosons that have transitioned into $m=-1$ fall back into
the BH; it sets the spectral width of the signal.  Equation~\eqref{eq:binary_h_plus_fourier}
describes a Lorentzian-shaped frequency-domain burst centered at the peak
frequency $f_p$ (Eq.~\eqref{eq:peak_freq}), with width $\sim |\Gamma|/\pi$.

The orbital chirp rate, derived from the leading-order gravitational-wave energy
loss of a circular binary with mass ratio $q \equiv M_c/M$, is
\begin{equation}
  \frac{\gamma}{\Omega_0^2} = \frac{96}{5}\,
    \frac{q}{(1+q)^{1/3}}\,(GM\Omega_0)^{5/3}\,.
  \label{eq:gamma_chirp}
\end{equation}

The adiabaticity parameter $z$ governs the degree of population transfer during
the resonance.  It is defined as
\begin{equation}
  z \equiv \frac{\eta^2}{|\Delta m|\,\gamma}\,,
  \label{eq:adiabaticity}
\end{equation}
where $\eta = \langle\psi_f|V_*(t,\bm{r})|\psi_i\rangle$ is the tidal mixing
amplitude between the $\{2,1,1\}$ and $\{2,1,-1\}$ states evaluated numerically
from the companion's gravitational potential $V_*$.  In the limit $z \ll 1$ the
resonance is non-adiabatic: the orbital frequency sweeps through the Bohr frequency
too rapidly for the levels to equilibrate, and only a fraction $\sim z$ of the
cloud population is transferred.  In the opposite limit $z \gg 1$ the resonance
is adiabatic and the populations invert almost completely.  In practice, however,
the final state $\{2,1,-1\}$ has a non-zero instability rate $|\Gamma|$ that
continuously drains bosons back into the BH, effectively truncating the transition
and limiting the maximum transferred population regardless of $z$.  This instability
is fully accounted for in Eq.~\eqref{eq:binary_h_plus_fourier}

As the binary sweeps through the resonance band, the GW signal duration is
\begin{equation}
  \Delta t = \frac{2|\Gamma|}{\gamma}(1 + 2z)\,.
  \label{eq:binary_signal_duration}
\end{equation}
The peak GW frequency is shifted from the carrier by the finite-$z$ correction,
\begin{equation}
  f_p = f_c - \frac{2z|\Gamma|}{\pi} = \frac{2 \Omega_0}{\pi} - \frac{2z|\Gamma|}{\pi}\,.
  \label{eq:peak_freq}
\end{equation}
For $z \ll 1$ the peak frequency approaches the orbital carrier, $f_p \to f_c$;
for $z \gg 1$ the signal is redshifted from the carrier by a term proportional
to $|\Gamma|$, quantifying the downward pull of the final-state decay.

\subsection{Parameter Space Constraints and Cavity-Response Diagnostic}
\label{subsec:param_space}
\label{sec:aq_constraints}  

\begin{figure}[t!]
\begin{center}
\includegraphics[width=0.7\textwidth,height=0.5\textwidth]{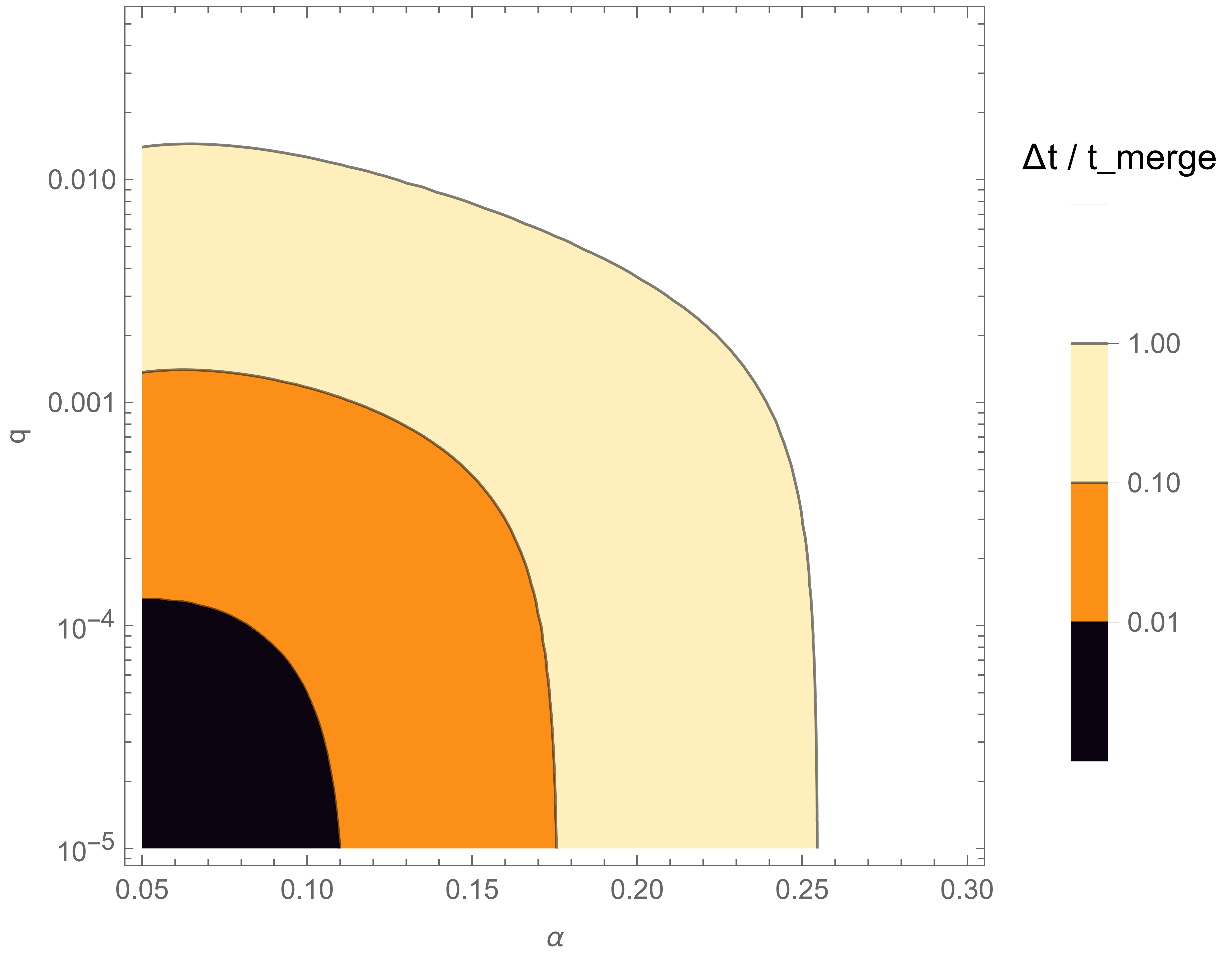}
\caption{Ratio $\Delta t/t_{\rm merge}$ (Eqs.~\eqref{eq:binary_signal_duration} and
  \eqref{eq:binary_merging_timescale}) in the $(\alpha, q)$ parameter plane for the
  $\{2,1,1\}\to\{2,1,-1\}$ transition, evaluated at the resonant orbital frequency
  $\Omega_0$.  Contours correspond to $\Delta t/t_{\rm merge} = 0.01$ (inner), $0.1$
  (middle), and $1$ (outer).  The white region, where $\Delta t/t_{\rm merge} > 1$,
  is excluded: the signal duration would exceed the remaining merger time, violating
  the linear-orbit approximation, and the binary would not be intact.  The allowed region is bounded by $\alpha \lesssim
  0.26$ and $q \lesssim 0.01$; throughout Secs.~\ref{subsec:pbh_detect}
  and~\ref{sec:pbh-rates} we adopt the benchmark values $\alpha = 0.14$ and
  $q = 10^{-4}$, which lie well within this allowed region.}
\label{fig:a-q constraints}
\end{center}
\end{figure}

Several physical requirements simultaneously constrain the allowed $(\alpha, q)$
parameter space.  The binary orbit is treated in the linear-chirp approximation
(Eq.~\eqref{eq:gamma_chirp}), which is valid only when the fractional frequency
change during the resonance is small: $\Delta\Omega/\Omega_0 = \gamma\,\Delta t/\Omega_0 \ll 1$.
The annihilation timescale of the initial $\{2,1,1\}$ state (Eq.~\eqref{eq:tau_ann})
must also exceed the resonance band crossing time $\Delta t$, otherwise the cloud is
depleted by annihilation before it can undergo the binary-driven transition.
In practice, both of these are less restrictive than the most stringent condition,
which requires the signal duration to be shorter than the binary merger
timescale:
\begin{equation}
  \Delta t < t_{\rm merge}\,,
  \label{eq:dt_lt_tmerge}
\end{equation}
where the Peters-formula merger time at orbital frequency $\Omega_0$ is
\begin{equation}
  t_{\rm merge} = \frac{5}{256}\,G^{-5/3}\,
    \frac{M}{(M\Omega_0)^{8/3}}\,\frac{(1+q)^{1/3}}{q}\,.
  \label{eq:binary_merging_timescale}
\end{equation}
The condition~\eqref{eq:dt_lt_tmerge} ensures that the resonant transition occurs
while the binary is still intact—a prerequisite for the waveform model of
Sec.~\ref{subsec:gab_waveform} to apply. Figure~\ref{fig:a-q constraints} displays the ratio $\Delta t/t_{\rm merge}$ in the
$(\alpha, q)$ plane, with contours at $\Delta t/t_{\rm merge} = 0.01, 0.1, 1$.  The
white region ($\Delta t/t_{\rm merge} > 1$) is excluded because the transition would
not complete before merger.  The figure shows that the allowed region is bounded by
$\alpha \lesssim 0.26$ and $q \lesssim 0.01$. We enforce these constraints in all
subsequent calculations in this section. 

For resonant cavities such as ADMX, in order to implement the steady-state sensitivity curves within Sec.~\ref{subsec:pbh_detect}, a further parameter space constraint is introduced by the requirement of a fully rung up cavity. A useful steady-state diagnostic is
whether the duration of the GW burst within the cavity bandwidth satisfies
\begin{equation}
  \Delta t \;\gtrsim\; \tau_{\rm ring} \;\equiv\; \frac{1}{\Delta f_{\rm band}}\,,
  \label{eq:ringup_condition}
\end{equation}
where $\Delta f_{\rm band} \simeq f/Q$ is the cavity bandwidth and $\tau_{\rm ring}$
is the ring-up time required for the cavity to reach its steady-state response.
A single cavity mode with resonance frequency $\omega_{\rm cav}$ and damping rate
$\gamma_{\text{cav}}=\omega_{\rm cav}/Q$ responds to a gravitational-wave strain $h(t)$ as a driven
oscillator,
\begin{equation}
\dot A(t)+\left[\frac{\gamma_{\text{cav}}}{2}+i\bigl(\omega_{\rm cav}-\omega_{\rm GW}(t)\bigr)\right]A(t)=\kappa h(t),
\label{eq:cavity_response_ode}
\end{equation}
where $A(t)$ is the electromagnetic mode amplitude and $\kappa$ encodes the magnetic
field, cavity geometry, and mode-overlap factor.  Equivalently,
\begin{equation}
A(t)=\kappa\int^t_{-\infty} dt'\,h(t')\exp\left[-\frac{\gamma_{\text{cav}}}{2}(t-t')-i\int_{t'}^t dt''\bigl(\omega_{\rm cav}-\omega_{\rm GW}(t'')\bigr)\right].
\label{eq:cavity_response_convolution}
\end{equation}
For a monochromatic signal with duration $\Delta t\gg \gamma_{\text{cav}}^{-1}$ this expression
reduces to the familiar steady-state response used in standard cavity sensitivity
estimates.  

For a chirping signal, which is the relevant case for this section as the binary sweeps through a resonant transition band that is much larger than the cavity bandwidth, the deposited energy is
controlled instead by the time, $t_{\rm dwell}\sim \Delta\omega_{\rm cav}/|\dot\omega_{\rm GW}|$, the waveform spends within the cavity bandwidth, and the detailed evolution of the phase-difference between the chirping GW and the cavity mode across the cavity bandwidth. The latter determines the decoherence time, $t_{\text{decoherence}}$, when the driving signal ceases to oscillate in-phase and coherent with the cavity mode to effectively deposit energy within the cavity, which limits the number of cavity mode oscillations that can be amplified by the driving force for a cavity ring up. For linear chirping signals $\omega(t) = \omega_{\rm cav} + \chi t $, an estimate for the decoherence time can be found as \cite{Schenk:2025cavity}:

\begin{equation}
\label{eq:time_decoherence_generic}
t_{\text{decoherence}} \sim \sqrt{\frac{\pi}{\chi}}
\end{equation}
which can also be considered as the coherence time interval $t_{\text{coherence}}$ that sets the maximum number of allowed oscillations, $N_{max}$, where the mode function amplitude grows.

Recall from Sec.~\ref{subsec:gab_waveform}, the chirping signal is:
$\omega_{GW}(t) = 4 \Omega_{0} + 4 \gamma t$.  At the $z \ll 1$ limit where $f_p \to f_c$, we have $4 \Omega_{0} + 4 \gamma t \sim \omega_{\rm cav} + \chi t$, such that $\chi = 4 \gamma$. The decoherence time for our considerations then becomes:

\begin{equation}
\label{eq:time_decoherence_binary}
t_{\text{decoherence}} \sim \sqrt{\frac{\pi}{4\gamma}}
\end{equation}
The criterion for determining whether a chirping signal is capable of satisfying the condition for applying the steady-state comparison is 
$\Delta t_{\rm eff}\gg \tau_{\rm ring}$, where $\Delta t_{\rm eff}\sim \min\{t_{\text{decoherence}}, t_{\text{dwell}} \}$, this ensures the GW signal is capable of fully ringing up the cavity within the allowed energy deposit time window. 

\begin{figure}[t!]
\begin{center}
\includegraphics[width=0.8\textwidth,height=0.5\textwidth]{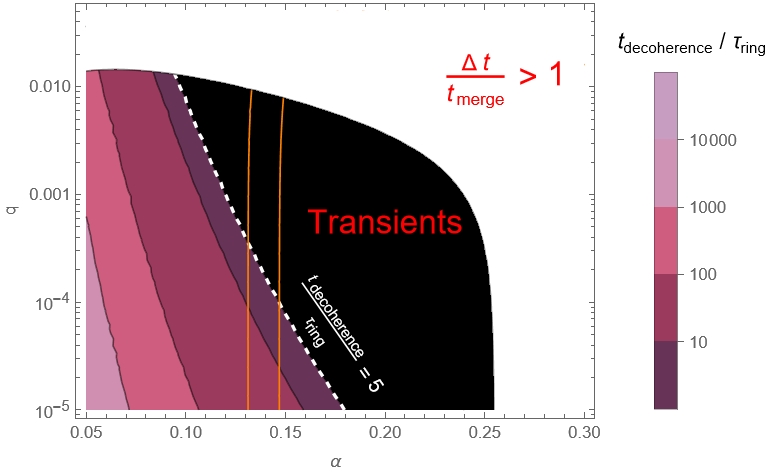}
\caption{Ratio of $\Delta t_{\text{eff}} / \tau_{\text{ring}}$ for the conservative response diagnostic with $\Delta t_{\rm eff}\sim t_{\text{decoherence}}$(Eq.~(\ref{eq:time_decoherence_binary})) within the allowed parameter space bounded by Fig.~\ref{fig:a-q constraints} for an ADMX-like cavity with $Q \sim 8\times10^4$, but without restricting the cavity resonance frequency to the ADMX hardware band contained between the orange solid curves. The white dash line corresponds to the boundary of $\Delta t_{\text{eff}} / \tau_{\text{ring}} = 5$, where binaries that produce smaller timescale ratios would correspond to transient signals incompatible with the steady-state approximation.}
\label{fig:dwelltimeconstraint}
\end{center}
\end{figure}

To provide an estimation on binary source parameters that are capable of ringing up the cavity, in Figure \ref{fig:dwelltimeconstraint} we plot the ratio of $\Delta t_{\text{eff}} / \tau_{\text{ring}}$ within the allowed parameter space defined by the binary merger time in Fig.~\ref{fig:a-q constraints} for an ADMX-like cavity with $Q \sim 8\times10^4$, but without restricting the cavity resonance frequency to the ADMX hardware band. Since  $t_{\text{decoherence}}$ (Eq. (\ref{eq:time_decoherence_binary})) provides a tighter constraint against $\tau_{\text{ring}}$ when compared to $t_{\text{dwell}}$ within the parameter space, we take $\Delta t_{\rm eff}\sim t_{\text{decoherence}}$. We identify GWs from the region where $t_{\text{decoherence}} / \tau_{\text{ring}} < 5$ as "Transiets", signals that would fail to ring up the cavity to apply the steady-state approximation.

However, we stress that Fig.~\ref{fig:dwelltimeconstraint} serves as an order-of-magnitude estimation, and should not be treated as a strict detection forecast. We also point out that generically, the ring-up condition should not be interpreted as a sharp detectability boundary, but only as a prerequisite for applying the steady-state sensitivity curves. Signals that fail to fully ring up the cavity could still deposit considerable amounts of energy to enable a measurable response. Nevertheless, for estimation purposes, in the remainder of the section we constraint ourselves within the region of Fig.~\ref{fig:dwelltimeconstraint} where the steady-state approximation remains valid.

\subsection{Detectability Prospects for PBH Gravitational-Atom Binaries}
\label{subsec:pbh_detect}

We now apply the binary-perturbation formalism to PBH systems and assess
detectability at ADMX.  To avoid overestimating the signal we adopt conservative
numerical inputs throughout: (i) relativistic instability rates from
Refs.~\cite{Hoof_2025, Hoof2025Code} for the final-state decay rate $\Gamma$, rather
than the leading-order analytic approximation of Eq.~\eqref{eq:SR_rate}; (ii) numerically
computed mixing amplitudes $\eta = \langle\psi_f|V_*(t,\bm{r})|\psi_i\rangle$; and
(iii) numerically computed cloud masses $M_{\rm cloud}$ from Refs.~\cite{Siemonsen:2022yyf, May:2024}.

\subsubsection{Peak-frequency map in the PBH mass range}

\begin{figure}[t!]
\begin{center}
\includegraphics[width=0.9\textwidth,height=0.55\textwidth]{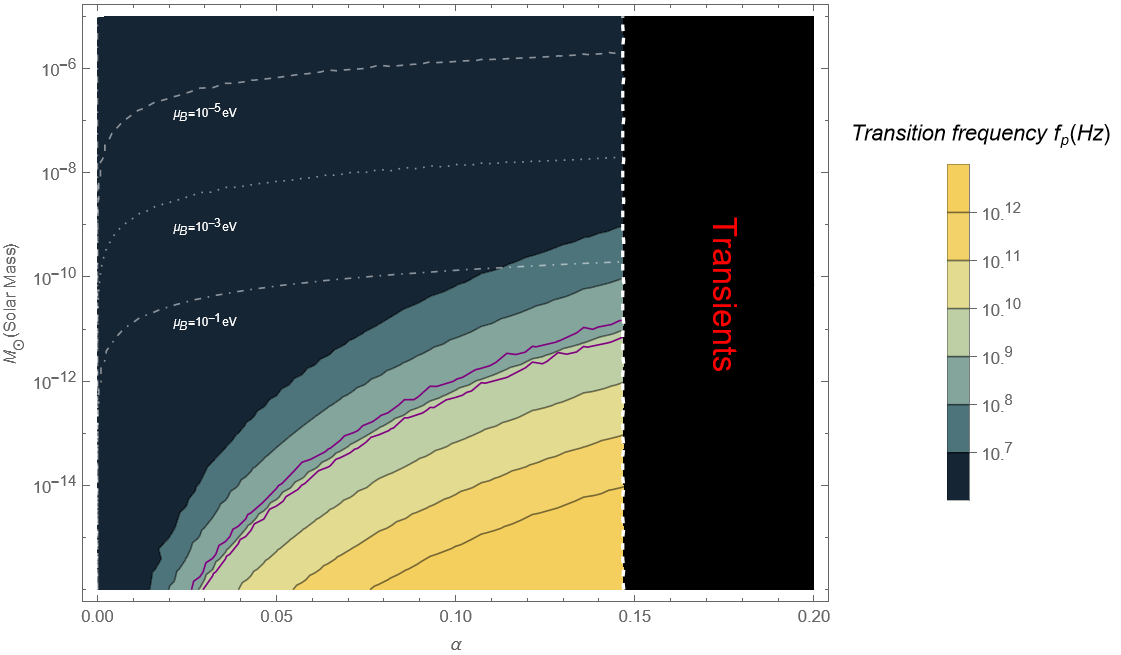}
\caption{Peak transition frequency $f_p$ (Eq.~\eqref{eq:peak_freq}) in the
  $M_{\rm BH}$--$\alpha$ parameter space for the $\{2,1,1\}\to\{2,1,-1\}$ hyperfine
  transition, for boson masses $\mu_b = 10^{-5}$\,eV (top), $10^{-3}$\,eV (middle),
  and $10^{-1}$\,eV (bottom), with $q = 10^{-4}$ fixed. The $\alpha$ value that defines the transient signal boundary shifts with different choices of $q$, as shown in Fig.~\ref{fig:dwelltimeconstraint}.  The PBH mass range cropped to
  $M_{\rm BH}\in[10^{-16},10^{-5}\,M_\odot]$ is shown, corresponding to the HF-band.  The ADMX Run-1 scan band
  (0.645--1.4\,GHz) is highlighted in purple; it is accessible for $M_{\rm BH}\sim
  10^{-11}$--$10^{-12}\,M_\odot$ and $\mu_b \sim 2.5$\,eV at $\alpha \sim 0.14$.
  Sources lying in this band are the primary targets of the detectability analysis
  in Sec.~\ref{subsec:pbh_detect}.}
\label{fig:fouier_peak_pbh}
\end{center}
\end{figure}

Figure~\ref{fig:fouier_peak_pbh} shows the peak GW frequency $f_p$
(Eq.~\eqref{eq:peak_freq}) across the $M_{\rm BH}$--$\alpha$ parameter space for
several values of the boson mass $\mu_b$, with the companion-to-host mass ratio
fixed at $q = 10^{-4}$.  We restrict to the PBH mass range $M_{\rm BH} \in
[10^{-16}, 10^1\,M_\odot]$ and exclude parameter combinations outside the
allowed region of Fig.~\ref{fig:a-q constraints}.  For the hyperfine transition
$\{2,1,1\}\to\{2,1,-1\}$, the superradiance condition $\alpha < m/2 = 0.5$ is
automatically satisfied within the constrained region.

The figure demonstrates that GW signals from binary-perturbed GA systems
naturally populate the ultra-high-frequency (UHF) band—including the GHz range
targeted by ADMX—\textit{if and only if} the host BH is of primordial origin.  The
ADMX Run-1 frequency scan range (0.645--1.4\,GHz, shown in purple) is intersected
by trajectories with $M_{\rm BH} \sim 10^{-11}$--$10^{-12}\,M_\odot$ and $\mu_b
\sim 2.5$\,eV at $\alpha \sim 0.14$. 

\subsubsection{Benchmark waveform}

\begin{figure}[h!]
  \centering
  \mbox{\hspace*{-0.2cm}\includegraphics[width=0.48\textwidth]{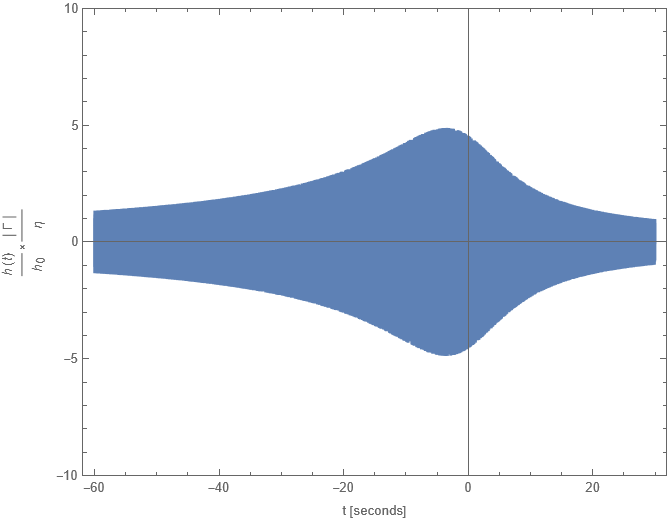}
  \hspace{0.005\textwidth}
  \quad\includegraphics[width=0.5\textwidth]{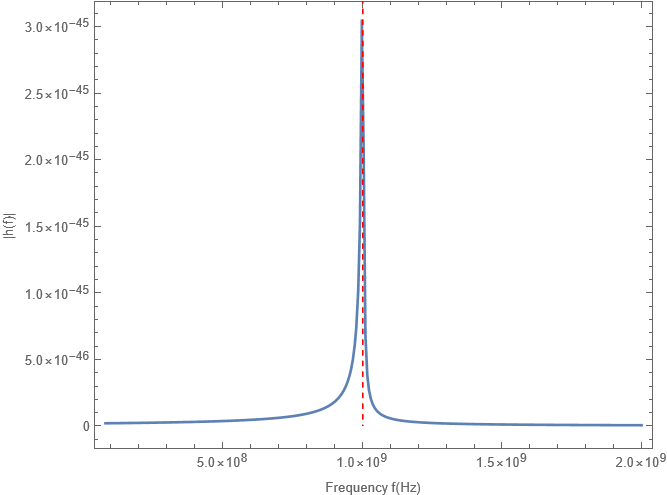}}
  \caption{Time-domain strain $h_+(t)$ (left) and frequency-domain amplitude
    $|\tilde{h}(f)|$ (right) for the benchmark binary GA system of
    Eq.~\eqref{eq:benchmark_params}: $\alpha = 0.14$, $M = 6.68\times10^{-12}\,M_\odot$,
    $q = 10^{-4}$, $r = 10\,\text{kpc}$.  In the time domain the signal is a
    short quasi-monochromatic burst with duration $\Delta t$ (Eq.~\eqref{eq:binary_signal_duration}).
    In the frequency domain the signal is a Lorentzian-shaped peak centered at the
    peak frequency $f_p = 1.001\,\text{GHz}$ (red dashed line), with spectral
    width $\sim |\Gamma|/\pi$.  Both panels are computed using
    Eqs.~\eqref{eq:binary_signal_h+} and \eqref{eq:binary_h_plus_fourier} with
    the numerical inputs described in Sec.~\ref{subsec:pbh_detect}.}
  \label{fig:pbh_h(t)_h(f)}
\end{figure}

Figure~\ref{fig:pbh_h(t)_h(f)} shows the time-domain strain $h_+(t)$ and its
Fourier transform $\tilde{h}(f)$ for the benchmark binary GA system with
\begin{equation}
  \alpha = 0.14,\quad
  M = 6.68\times10^{-12}\,M_\odot,\quad
  q = 10^{-4},\quad
  r = 10\,\text{kpc}.
  \label{eq:benchmark_params}
\end{equation}
These parameters lie within the allowed region of Fig.~\ref{fig:a-q constraints} and Fig.~\ref{fig:dwelltimeconstraint},
place the peak frequency $f_p = 1.001\,\text{GHz}$ within the ADMX band, and
correspond to the event rates computed in Sec.~\ref{sec:pbh-rates}.  The resulting
signal is a quasi-monochromatic GW burst.

\subsubsection{ADMX sensitivity comparison}
\label{subsec: ADMX_sensitivity}

The minimum GW strain detectable by a resonant cavity is~\cite{BerlinHF}
\begin{align}
  h_{\rm min} = 3\times10^{-22}
    &\left(\frac{0.1}{\eta_n}\right)
    \left(\frac{8\,\text{T}}{|\mathbf{B}|}\right)
    \left(\frac{0.1\,\text{m}^3}{V_{\rm cav}}\right)^{\!5/6}
    \left(\frac{10^5}{Q}\right)^{\!1/2}
    \nonumber\\
    &\times
    \left(\frac{T_{\rm sys}}{1\,\text{K}}\right)^{\!1/2}
    \left(\frac{1\,\text{GHz}}{f}\right)^{\!3/2}
    \left(\frac{\Delta f}{10\,\text{kHz}}\right)^{\!1/4}
    \left(\frac{1\,\text{min}}{t_{int}}\right)^{\!1/4},
  \label{eq:ADMX_sensitivity}
\end{align}
where $\eta_n$ is the cavity--GW coupling coefficient, $|\mathbf{B}|$ is the
axial magnetic field, $V_{\rm cav}$ is the cavity volume, $Q$ is the quality
factor, $T_{\rm sys}$ is the system noise temperature, $f$ is the resonant
frequency, $\Delta f \simeq f/Q$ is the cavity bandwidth, and $t_{int}$ is the measurement integration time. For ADMX Run 1 the
relevant parameters are $\eta_n \sim 0.1$, $|\mathbf{B}| = 0.75\,\text{T}$,
$V_{\rm cav} = 136\,\text{L}$, $Q \sim 8\times10^4$, $T_{\rm sys} = 0.6\,\text{K}$,
, $f \in [0.65, 1.4]\,\text{GHz}$, and $t_{int} = 2\,\text{min}$~\cite{BerlinHF}, giving a typical threshold
$h_{\rm min} \sim 10^{-22}$ within the ADMX band.  The GW characteristic strain
for binary-induced transitions is $h_c(f) = 2f|\tilde{h}(f)|$ with $\tilde{h}(f)$
given by Eq.~\eqref{eq:binary_h_plus_fourier}, Fig. \ref{fig:pbh_charateristic} shows the characteristic strain for the benchmark parameters of Eq.~\eqref{eq:benchmark_params}.

\begin{figure}[t!]
\begin{center}
\includegraphics[width=0.9\textwidth,height=0.6\textwidth]{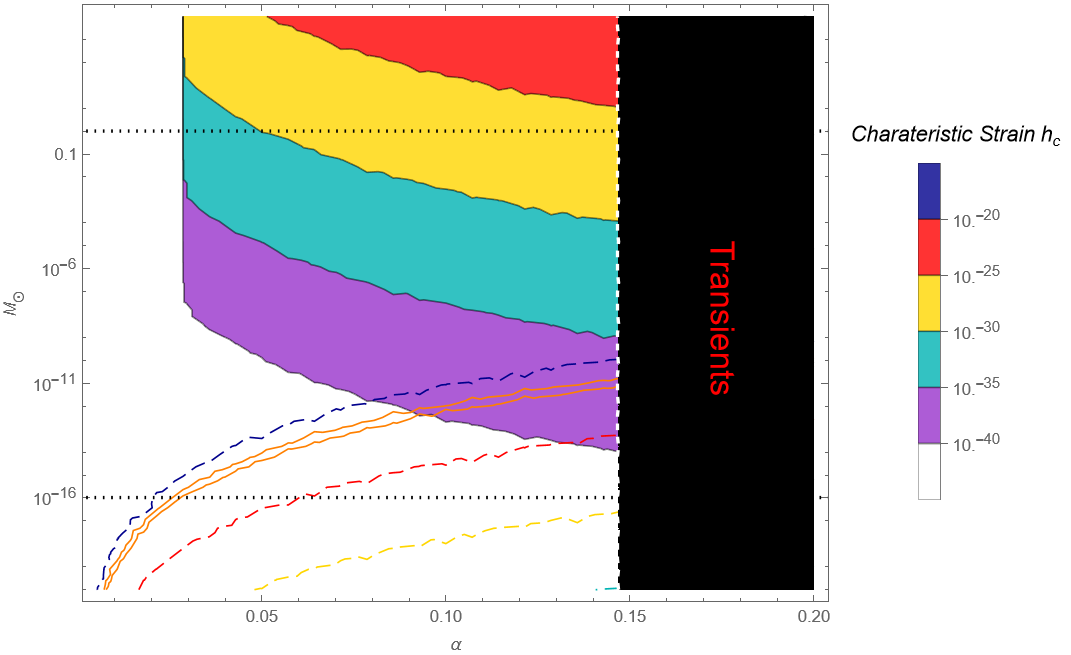}
\caption{GW characteristic strain $h_c$ (filled contours) and ADMX minimum
  detectable strain $h_{\rm min}$ (dashed contours, Eq.~\eqref{eq:ADMX_sensitivity})
  for binary GA systems in the $M_{\rm BH}$--$\alpha$ plane, with $q = 10^{-4}$ and
  $r = 10\,\text{kpc}$.  Filled contour levels span $h_c \in [10^{-40}, 10^{-20}]$;
  dashed contour levels show the same decades of $h_{\rm min}$, computed without
  restricting $f$ to the ADMX hardware band.  The orange solid curves overlay the
  ADMX frequency-band boundaries from Fig.~\ref{fig:fouier_peak_pbh}, delimiting
  the region where ADMX can physically respond; within this region $h_{\rm min}
  \sim 10^{-22}$.  The PBH mass range $10^{-16}\,M_\odot \lesssim M_{\rm PBH}
  \lesssim 1\,M_\odot$ is bounded by the black dotted lines.  No $h_c$ contour
  overlaps a $h_{\rm min}$ contour of equal magnitude within the PBH mass range
  at this distance, demonstrating that the signal falls orders of magnitude below
  the ADMX detection threshold.}
\label{fig:ADMX_GWStrain_Compare}
\end{center}
\end{figure}

Figure~\ref{fig:ADMX_GWStrain_Compare} displays both $h_c$ (filled contours) and
$h_{\rm min}$ (dashed curves) across the $M_{\rm BH}$--$\alpha$ plane at $q = 10^{-4}$
and $r = 10\,\text{kpc}$.  The dashed contours of $h_{\rm min}$ are evaluated
at fixed values of the ADMX sensitivity parameter (Eq.~\eqref{eq:ADMX_sensitivity})
but without restricting $f$ to the ADMX hardware band; the orange solid curves
then overlay the ADMX frequency range from Fig.~\ref{fig:fouier_peak_pbh} to
delineate the region where the ADMX hardware could physically respond, during Run1 data, in turn corresponding
to $h_{\rm min} \sim 10^{-22}$.  The PBH mass range of primary interest,
$10^{-16}\,M_\odot \lesssim M_{\rm PBH} \lesssim 1\,M_\odot$, is bounded by the
black dotted lines.

For the benchmark parameters at $r = 10\,\text{kpc}$, the characteristic strain
$h_c$ falls \textit{several orders of magnitude} below the ADMX detection threshold
throughout the accessible parameter space.  No overlap between the $h_c$ and
$h_{\rm min}$ contours of equal magnitude is found within the PBH mass range,
confirming that ADMX as currently configured cannot detect the binary-induced burst
at astrophysically motivated distances.

The gap between $h_c$ and the ADMX threshold persists even when one optimizes
over the remaining free parameters.  Increasing the mass ratio $q$ toward its
upper bound $q \lesssim 0.01$ (Sec.~\ref{subsec:param_space}) enhances $h_c$ only
modestly, with the most direct path to improved sensitivity is reducing the
source distance $r$: from Eq.~\eqref{eq:h0_factor}, $h_0 \propto 1/r$, so ADMX
threshold strains are reached only at $r \lesssim 1\,\text{AU}$.  Such distances
are strongly disfavored by merger-rate arguments, as we demonstrate quantitatively
in Sec.~\ref{sec:pbh-rates}.

We reiterate that this comparison is intended as an order-of-magnitude benchmark rather
than a complete chirping/transient-search forecast, as discussed in Sec.~\ref{subsec:param_space}.  For the isolated annihilation and
level-transition signals of Sec.~\ref{sec:isolated-GA}, whose lifetimes are vastly
longer than any cavity response time, the steady-state sensitivity curve is the
appropriate limiting case.  For the binary-induced signals considered here, a rigorous
analysis would require filtering the time-domain waveform through the cavity transfer
function, or equivalently computing the total deposited electromagnetic energy in the
relevant modes.  Recent studies of electromagnetic cavities exposed to non-monochromatic
high-frequency gravitational waves show that the loss of sensitivity for short or
chirping signals need not be captured by a step-function ring-up criterion
alone~\cite{Schenk:2025cavity,Blas:2026multimode,Amaral:2026gravnet}.  Regardless,
our main conclusion for binary-induced gravitational-atom bursts is insensitive to
this refinement: at astrophysically plausible distances the predicted strain lies
many orders of magnitude below the current ADMX-equivalent sensitivity, and the
event-rate argument requires detectable sources to occur at unrealistically small
distances.  A more complete detector-response calculation may change the precise
numerical reach, but it cannot remove the combined strain and rate deficits identified
here.

\subsection{Doppler Frequency Shift for Solar-System Flybys}
\label{subsec:doppler}

The only scenario in which a PBH binary could approach within $\sim 1\,\text{AU}$
is a transient flyby through the solar system.  In this case the relative velocity
between Earth and the binary imprints a Doppler shift on the observed GW frequency.
We estimate the maximum frequency shift by setting the flyby velocity equal to the
local dark-matter virial velocity, $v_{\rm DM} \approx 200\,\text{km\,s}^{-1}$:
\begin{equation}
  \delta f = \frac{v_{\rm DM}}{c}\,f_p
  \;\approx\; 6.7\times10^{-4}\times f_p\,.
  \label{eq:doppler_shift}
\end{equation}
For the benchmark peak frequency $f_p = 1.001\,\text{GHz}$, this gives $\delta f
\approx 670\,\text{kHz}$.  Although this shift exceeds the instantaneous cavity
bandwidth of ADMX ($\Delta f_{\rm band} \simeq f/Q \approx 13\,\text{kHz}$) and
would therefore displace the signal out of resonance during a naive fixed-frequency
scan, it remains a small fraction of the full ADMX Run-1 frequency scan range
$(1.4 - 0.645)\,\text{GHz} = 755\,\text{MHz}$.  A flyby source would therefore
be visible in some portion of the scan range, and a matched-filter search that
accounts for a $\pm 1\,\text{MHz}$ Doppler drift would recover the full
signal duration.

We note, however, that the detectability problem identified in
Sec.~\ref{subsec:pbh_detect} is one of \textit{strain} sensitivity, not frequency
coverage: even at $r = 1\,\text{AU}$ the strain falls roughly six orders of
magnitude below the ADMX threshold. The Doppler correction is therefore not the obstacle to detection.

\section{Prospects for Observation at High-Frequency Gravitational-Wave Detectors}
\label{sec:pbh-rates}

The detectability analysis of Sec.~\ref{subsec:pbh_detect} shows that the
characteristic strain of the binary-induced burst falls several orders of magnitude
below the ADMX threshold at $r = 10\,\text{kpc}$, requiring systems which can produce binary-induced level transitions to exist at much closer distances, potentially $\ll1$ AU. Whether such close encounters occur at any
appreciable rate depends on the abundance and spatial distribution of PBHs, and on
the statistics of PBH binary formation.  Because the orbital separations that drive
the $\{2,1,1\}\to\{2,1,-1\}$ resonant transition occur immediately prior to merger
(Sec.~\ref{subsec:param_space}), the question reduces to the local PBH merger rate
filtered by the conditions on mass ratio and total binary mass identified in
Sec.~\ref{subsec:pbh_detect}: $q \lesssim 10^{-4}$ and $M=6.68\times10^{-12}M_{\odot}$.

In Ref.~\cite{Raidal:2024bmm}, merger rates were computed for PBHs formed in the
early universe across four binary-formation channels and for generic mass functions.
We focus on the \textit{early two-body channel}, in which two neighboring PBHs form
a gravitationally bound pair when their mutual attraction overcomes the Hubble flow.
This channel is conceptually the cleanest, yields the most direct mapping to the
$(M, q)$ space, and dominates the total merger rate at low $f_{\rm PBH}$, where $f_{\rm PBH}$ is the PBH fraction of the dark-matter energy density.
The competing \textit{early three-body channel}, which can become dominant at large $f_{\rm PBH}$,
requires a treatment of N-body perturbations to the initial binary orbit and carries
larger theoretical uncertainties~\cite{Raidal:2024bmm}; we discuss its potential
impact briefly in Sec.~\ref{subsec:clustering} below.

\subsection{Merger Rate from the Early Two-Body Formation Channel}
\label{subsec:E2_rate}

The differential merger rate density per unit logarithmic mass for the early
two-body channel is given in \cite{Raidal:2024bmm} as:
\begin{equation}
  \frac{{\rm d}R_{E2}}{{\rm d}\ln m_1\,{\rm d}\ln m_2}
  \approx
  \frac{1.6\times10^6}{\rm Gpc^3\,yr}\;
  f_{\rm PBH}^{53/37}\,
  \eta_{\rm PBH}^{-34/37}\,
  \!\left(\frac{M_{\rm tot}}{M_\odot}\right)^{\!-32/37}
  \!\left(\frac{t}{t_0}\right)^{\!-34/37}
  S_L\,S_E\,\psi(m_1)\,\psi(m_2)\,,
  \label{eq:E2_rate}
\end{equation}
where 
$m_{1}$ and $m_{2}$ are the masses of the two black holes in the binary, $\eta_{\rm PBH} \equiv \langle m\rangle^2/\langle m^2\rangle$ is the inverse
normalized second moment of the mass function $\psi(m)$ (not to be confused with
the tidal mixing amplitude $\eta$ of Eq.~\eqref{eq:adiabaticity}), $t$ is the merger time, and $t_0$ is the
age of the universe.  The suppression factor
\begin{equation}
  S_L \approx \min\!\left\{
    1,\;
    0.01\!\left[\!\left(\frac{t}{t_0}\right)^{\!0.44}\!f_{\rm PBH}\right]^{\!-0.65}
    \exp\!\left(0.03\ln^2\!\left[\left(\frac{t}{t_0}\right)^{\!0.44}\!f_{\rm PBH}\right]\right)
  \right\}
  \label{eq:SL}
\end{equation}
accounts for the disruption of PBH binaries by late-time large-scale structure, and
\begin{equation}
  S_E \approx
  \frac{\sqrt{\pi}\,(5/6)^{21/74}}{\Gamma(29/37)}
  \left[\frac{\langle m^2\rangle/\langle m\rangle^2}{\bar{N}(y) + C}
  + \frac{\sigma_M^2}{f_{\rm PBH}^2}\right]^{\!-21/74}
  e^{-\bar{N}(y)}
  \label{eq:SE}
\end{equation} where

\begin{equation}
C = f_{\rm PBH}^2 \, \frac{\langle m^2 \rangle / \langle m \rangle^2}{\sigma_M^2}
\left\{
\left[
\frac{\Gamma(29/37)}{\sqrt{\pi}} \,
U\!\left(\frac{21}{74}, \frac{1}{2}, \frac{5 f_{\rm PBH}^2}{6 \sigma_M^2}\right)
\right]^{-74/21}
- 1
\right\}^{-1}
\end{equation} captures the suppression from Poisson fluctuations in the number of neighboring
PBHs within the relevant comoving volume.  Here $\bar{N}(y)$ is the mean number
of PBH neighbors within comoving separation $y$, and $\sigma_M^2/f_{\rm PBH}^2$ is the variance of the
mass-weighted density contrast~\cite{Raidal:2024bmm}.

The rate in Eq.~\eqref{eq:E2_rate} is evaluated at $t = t_0$ (i.e., for mergers
occurring today), which gives the present-day merger rate density $R_{E2}(t_0)$
relevant for ADMX observations.  We integrate over $(m_1, m_2)$ via Monte-Carlo
to obtain the portion of the rate associated with binaries in any chosen
$(m_1, q)$ bin, using the relations $q = m_2/m_1$ (with $m_2 < m_1$).

\subsection{Dichromatic Mass Function and Event Rates}
\label{subsec:dichromatic}

To evaluate the merger rate for our benchmark system ($m_1 = 6.68\times10^{-12}\,M_\odot$,
$m_2 = 6.68\times10^{-16}\,M_\odot$, so $q = 10^{-4}$); we examine two mass functions capable of producing large numbers of mixed-mass binary systems: a dichromatic mass function
\begin{equation}
  \psi_{\rm di}(m) = f_1\,\delta(m - m_1) + f_2\,\delta(m - m_2)\,,
  \quad f_1 + f_2 = 1\,,
  \label{eq:dichromatic_mf}
\end{equation}
which places all PBH mass at two discrete values, and a double-lognormal mass function

\begin{equation}
  \psi_{\rm dl}(m) = 
  \sum_{i=1}^{2}
  \frac{f_i}{\sqrt{2\pi}\,\sigma\, m}
  \exp\!\left[-\frac{\ln^2(m/m_i)}{2\sigma^2}\right],
  \quad f_1 + f_2 = 1\,.
  \label{eq:doublelognormal_mf}
\end{equation}

Under Eq.~\eqref{eq:dichromatic_mf} exactly three binary configurations are possible: equal-mass pairs at $m_1$, equal-mass pairs at $m_2$, and unequal-mass
pairs $(m_1, m_2)$ with $q = 10^{-4}$. Under Eq.~\eqref{eq:doublelognormal_mf}, there is greater allowed variation in $m_1$ and $q$, but most systems obtain parameters close to the previous three scenarios. As $\sigma\rightarrow0$, the double lognormal case should reduce to the dichromatic case. 

The resulting present-day merger rate densities, computed by Monte-Carlo integration
of Eq.~\eqref{eq:E2_rate} with $f_{\rm PBH} = 1$ (all dark matter in PBHs, a
conservative upper bound on the rate), are summarized in Table~\ref{tab:merger_rates}.
The equal-mass $(m_2, m_2)$ configuration dominates at $R_{E2} = 4.9\times10^{17}\,
\text{Gpc}^{-3}\,\text{yr}^{-1}$.  The unequal-mass $(m_1, m_2)$ configuration yields $R_{E2}\approx2.3\times10^{17}\,\text{Gpc}^{-3}\,\text{yr}^{-1}$, and equal-mass $(m_1, m_1)$ configurations yield $R_{E2} \approx 3.2\times10^{12}\,
\text{Gpc}^{-3}\,\text{yr}^{-1}$.  Only the $(m_1, m_2)$ binaries satisfy the parameter space constraints
  $q \lesssim 10^{-2}$ bounded by Fig.~\ref{fig:a-q constraints}; this is the rate we use for the observation
probability estimate.

\begin{table}[t!]
\centering
\renewcommand{\arraystretch}{1.4}
\begin{tabular}{|lccc|}
\hline
Binary type & $q$ & $R_{E2}$ [Gpc$^{-3}$\,yr$^{-1}$] & ADMX-relevant? \\
\hline
$(m_1, m_1)$: $6.68\times10^{-12}, 6.68\times10^{-12}\,M_\odot$ & 1.0 &
  $3.2\times10^{12}$ & No \\
$(m_2, m_2)$: $6.68\times10^{-16}, 6.68\times10^{-16}\,M_\odot$ & 1.0 &
  $4.9\times10^{17}$ & No \\
$(m_1, m_2)$: $6.68\times10^{-12}, 6.68\times10^{-16}\,M_\odot$ & $10^{-4}$ &
  $2.3\times10^{17}$ & \textbf{Yes} \\
\hline
\end{tabular}
\caption{Present-day merger rate densities from the early two-body channel
  (Eq.~\eqref{eq:E2_rate}) for the dichromatic mass function
  (Eq.~\eqref{eq:dichromatic_mf}) with $m_1 = 6.68\times10^{-12}\,M_\odot$,
  $m_2 = 6.68\times10^{-16}\,M_\odot$, and $f_{\rm PBH} = 1$.
  Only $(m_1, m_2)$ binaries satisfy the parameter space constraints
  $q \lesssim 10^{-2}$ in Fig.~\ref{fig:a-q constraints}.}
\label{tab:merger_rates}
\end{table}

\subsection{Characteristic Detection Distance and the Strain Gap}
\label{subsec:clustering}

Given a volumetric merger rate $R_{E2}$ [Gpc$^{-3}$\,yr$^{-1}$], the characteristic
distance within which one event per year is expected is
\begin{equation}
  d_{1\,\rm yr} = \left(\frac{3}{4\pi R_{E2}}\right)^{\!1/3},
  \label{eq:d1yr}
\end{equation}
obtained by setting the expected number of events in a sphere of radius $d$ equal
to unity: $R_{E2} \cdot \tfrac{4}{3}\pi d^3 = 1$.  For the ADMX-relevant unequal-mass
$(m_1, m_2)$ rate $R_{E2} = 2.3\times10^{17}\,\text{Gpc}^{-3}\,\text{yr}^{-1}$,
Eq.~\eqref{eq:d1yr} gives
\begin{equation}
  d_{1\,\rm yr}^{(q=10^{-4})} \approx 1\,\text{kpc}\,.
\end{equation}
At this distance, the characteristic strain from Sec.~\ref{subsec:pbh_detect} ($h_{c}\sim10^{-37}$) falls
$\mathcal{O}(10^{15})$ times below the ADMX sensitivity threshold $h_{\rm min} \sim
10^{-22}$. In order for this deficit to be made up, a source would need to be placed within approximately 100 kilometers of ADMX, which is wholly unrealistic both due to the extremely low event rates corresponding to mergers at this distance and the complication that systems within this distance to ADMX would likely be strongly perturbed by the presence of the Earth.

\begin{figure}[t!]
\centering

\begin{minipage}{0.48\linewidth}
  \centering
  \begin{overpic}[width=\linewidth]{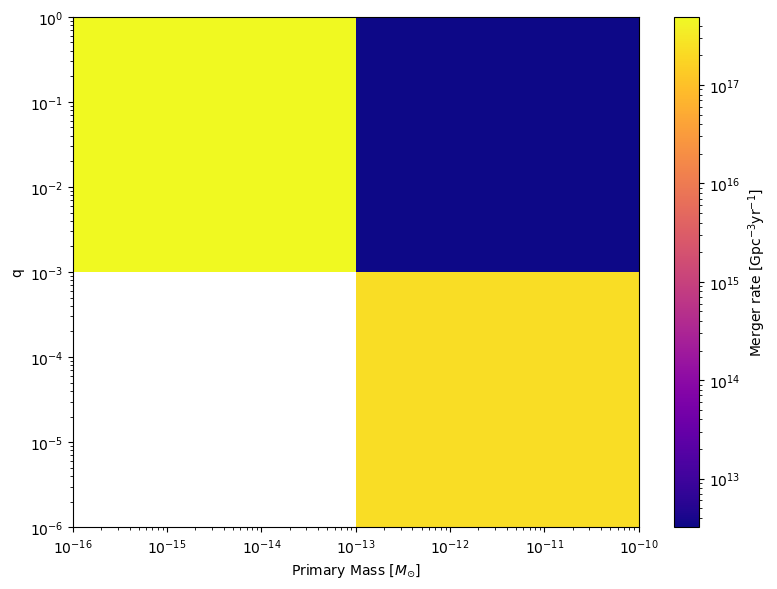}
    \put(5,80){\textbf{(a)}}
  \end{overpic}
\end{minipage}
\hfill
\begin{minipage}{0.48\linewidth}
  \centering
  \begin{overpic}[width=\linewidth]{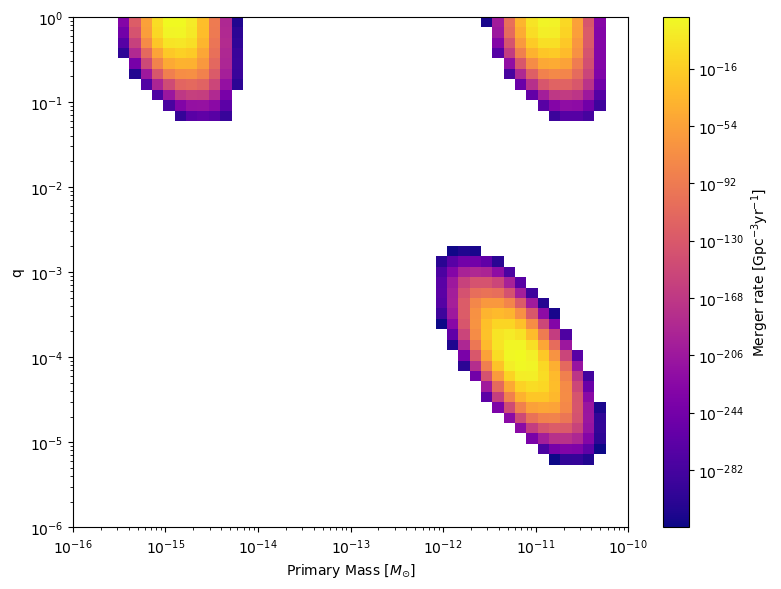}
    \put(5,80){\textbf{(b)}}
  \end{overpic}
\end{minipage}

\vspace{2.0em}

\begin{minipage}{0.48\linewidth}
  \centering
  \begin{overpic}[width=\linewidth]{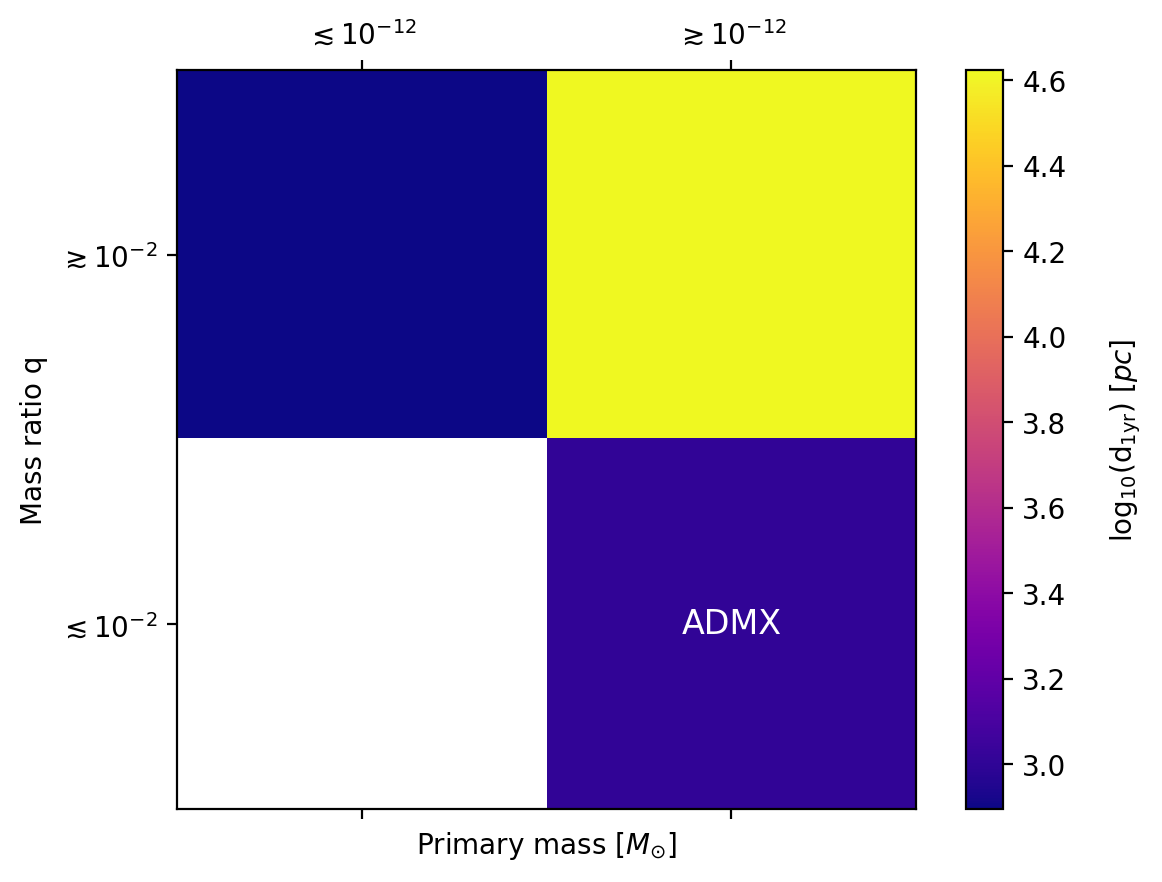}
    \put(5,80){\textbf{(c)}}
  \end{overpic}
\end{minipage}
\hfill
\begin{minipage}{0.48\linewidth}
  \centering
  \begin{overpic}[width=\linewidth]{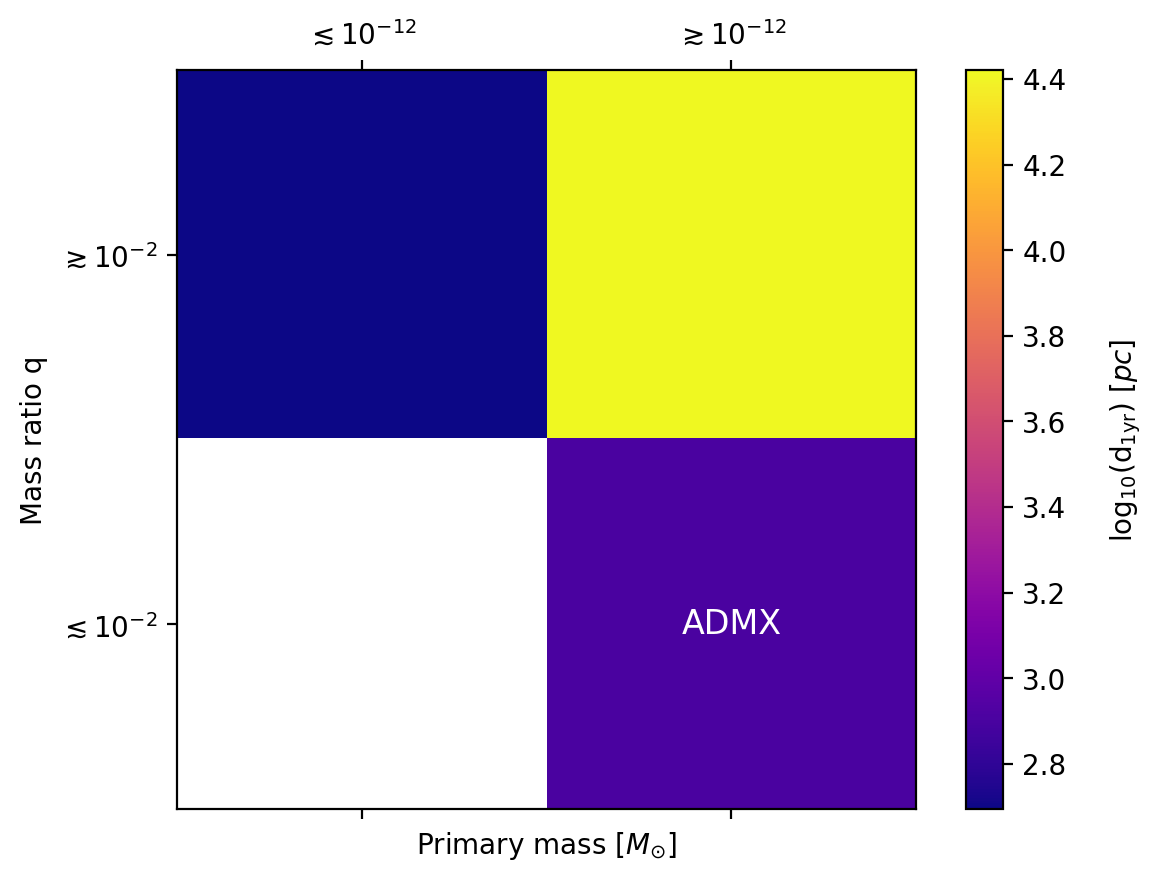}
    \put(5,80){\textbf{(d)}}
  \end{overpic}
\end{minipage}

\caption{(a-b) Merger rates in $M_{tot}-q$ space for dichromatic (a) and double lognormal (b) PBH mass functions with $m_1 = 6.68\times10^{-12}\,M_\odot$, $m_2 = 6.68\times10^{-16}\,M_\odot$, and $f_{1}=f_{2}=0.5$ for both functions and $\sigma=0.05$ for the double lognormal. 
(c-d) Characteristic distances $d_{1\,\rm yr}$ (Eq.~\eqref{eq:d1yr}) at which one merger per year is expected on average, for the dichromatic mass function (c) and for the double lognormal (d), subject to the binaries falling within a general range of benchmark parameters outlined in Sec .\ref{sec:binary-GA}.
For the dichromatic case, the three binary types correspond to those shown in Table~\ref{tab:merger_rates}. 
The ADMX-relevant $(m_1, m_2)$ binaries (labeled ``ADMX'') merge at appreciable rates only at $d \gtrsim 1\,\text{kpc}$ for both mass functions, whereas ADMX requires $r \ll 1\,\text{AU}$ to reach the detection threshold. 
The gap spans roughly fifteen orders of magnitude in strain.}
\label{fig:event_distance}
\end{figure}

Figure~\ref{fig:event_distance} summarizes the characteristic distances $d_{1\,\rm yr}$
for all three binary classes in the dichromatic model.  We find that a lognormal mass function centered on the same benchmark masses produces very similar results.

\paragraph{Effect of PBH clustering.}
Local overdensities of PBHs increase the effective merger rate by a factor $\delta$
relative to the homogeneous prediction, which rescales all merger rates in
Eq.~\eqref{eq:E2_rate} by $\delta$~\cite{Raidal:2024bmm}.  Since the event rate
scales as $R \propto \delta$, the characteristic distance scales as $d_{1\,\rm yr}
\propto \delta^{-1/3}$.  The most optimistic clustering estimates in the literature
give $\delta \lesssim 10^5$~\cite{Raidal:2024bmm}, which reduces $d_{1\,\rm yr}$
by a factor of $(10^5)^{1/3} \approx 46$:
\begin{equation}
  d_{1\,\rm yr}^{(q=10^{-4},\ \delta = 10^5)} \approx \frac{1\,\text{kpc}}{46} \approx 22\,\text{pc}.
\end{equation}
Even with this maximally optimistic clustering, the strain deficit from systems that merge at appreciable rates is improved by only one to two orders of magnitude—far from enough to make up the original $\mathcal{O}(10^{15})$ gap between the binary-induced level transition signal and the ADMX strain sensitivity. 

\paragraph{Effect of 3-Body Channel.} The prior analysis relies on estimating the merger rate of PBH systems generated via the ``early 2-body'' channel, where PBH binaries form shortly after PBH formation due to mutual gravitational attraction between close pairs. However, as noted in \cite{Raidal:2024bmm}, binaries can also form via encounters between early 2-body channels and isolated PBHs in what is known as the ``3-body channel''. These encounters lead to a population of PBH binaries with different orbital properties, which can increase their merger rates, especially when $f_{\rm PBH}$ approaches $1.0$. We find that, while this \textit{can} increase the PBH merger rate for the types of systems we are interested in, the increase in merger rate is typically not more than a few orders of magnitude, once again falling far below the level of enhancement we would need to bridge the detectability gap.

\subsection{Implications for Future Detectors}
\label{subsec:future_detectors}

The analysis above identifies two independent, multiplicative obstacles to
detecting binary-driven GA transitions at present-day cavity experiments:
\begin{enumerate}
  \item \textit{Strain deficit}: At the natural merger-rate distance $d_{1\,\rm yr}
    \approx 1\,\text{kpc}$, the signal strain is $\sim 10^{15}$ times below the ADMX
    threshold. A factor-of-$10^{15}$ improvement in strain sensitivity would be
    required from this factor alone.
  \item \textit{Rate deficit}: Reducing the merger distance to a few hundred kilometers while maintaining an appreciate merger rate requires an overall rate enhancement of $(1\,\text{kpc}/100\,\text{km})^3 \sim 10^{38}$ over
    the homogeneous prediction—far beyond what any clustering model can provide.
\end{enumerate}
These obstacles together imply that a qualitative change in experimental approach,
rather than incremental improvements to existing cavities, is necessary for binary-GA
signals to become observable.  The specific detector properties that would help most
are: (i) improved strain sensitivity, achievable in principle through larger effective
volume, stronger magnetic fields, lower system temperature, or enhanced mode overlap;
(ii) a response model and readout strategy optimized for transient and frequency-drifting
sources, rather than only for steady-state monochromatic signals; and (iii) broader
frequency coverage, especially at sub-GHz frequencies where heavier PBHs produce
signals outside the current ADMX scan range.  Recent proposals using deposited-energy
calculations for non-monochromatic cavity signals, multimode cavities that can track
frequency drift, and geographically separated detector networks provide promising
directions for going beyond the single-mode steady-state approximation used as our
baseline comparison~\cite{Schenk:2025cavity,Blas:2026multimode,Amaral:2026gravnet}.

The isolated-GA annihilation and level-transition signals studied in
Sec.~\ref{sec:isolated-GA} are not subject to the same rate limitations, since they
arise from persistent, continuous emission rather than merger events.  Their
detectability is governed instead by the peak strain at a given distance, with
$h_{0,\rm ann}^{\rm peak} \sim 10^{-22}$ at 1\,kpc (Eq.~\eqref{eq:hpeak_ann}).
For sources within $\sim 1\,\text{kpc}$—well within the Milky Way PBH halo—the
annihilation signal of an isolated GA is a more promising target for current
ADMX-band experiments than the binary-induced burst, and we highlight it as the
primary observational target for near-term searches.

\section{Discussion and Conclusions}
\label{sec:conclusions}

We have presented a unified treatment of high-frequency gravitational-wave emission
from superradiant boson clouds around primordial black holes, covering both isolated
gravitational atoms and binary-perturbed systems.  In this section we synthesize
the main results, place them in the context of the broader superradiance and
high-frequency GW literature, discuss the principal caveats and uncertainties, and
identify the most promising directions for future work.

\subsection{Summary and Key Conclusions}

\paragraph{Parameter space and Regge trajectories (Sec.~\ref{sec:Regge Trajectories}).}
The gravitational fine-structure constant $\alpha = GM_{\rm BH}\mu$ determines both
the efficiency of superradiant cloud growth and the frequency of the emitted GWs.
For efficient superradiance ($\alpha \in [0.05, 0.5]$) and GW emission in the
MHz--GHz band, the superradiance condition selects boson masses $\mu \sim
10^{-8}$--$10^{-5}\,\text{eV}$ and, through $M_{\rm BH} = \alpha/(G\mu)$, host BH
masses $M_{\rm BH} \sim 10^{-6}$--$10^{-3}\,M_\odot$—precisely the primordial BH mass
window (Table~1).  The Regge trajectories make this mapping explicit: each allowed
$(M_{\rm BH}, \mu, m)$ configuration traces a curve in the $(a_*, M_{\rm BH})$ plane
along which the cloud grows, extracts angular momentum, and emits narrowband GWs.
The maximum cloud mass fraction of $10.8\%$, obtained numerically for $\alpha \approx
0.2$ and $a_* \to 1$, is in close agreement with the analytic upper bound of
Ref.~\cite{Tomaselli:2024}.

\paragraph{Isolated gravitational atom (Sec.~\ref{sec:isolated-GA}).}
For an isolated GA we computed time-domain waveforms for both the level-transition
and boson-annihilation channels, together with analytic frequency-domain templates
(Appendix~\ref{sec:templates}).  The key quantitative results are:
\begin{itemize}
  \item \textit{Annihilation}: peak strain $h_{0,\rm ann}^{\rm peak} \simeq 10^{-21}$
    at $r = 145\,\text{au}$, with emission frequency $\omega_{\rm ann} \approx
    833.09\,\text{MHz}$ for the benchmark system ($\mu_a = 2\times10^{-6}\,\text{eV}$,
    $M_{\rm BH} = 1.7\times10^{-5}\,M_\odot$, $\alpha = 0.21$).  The signal is
    effectively a monochromatic steady source with a decay timescale $\tau_{\rm ann}
    \sim 10^{25}$--$10^{60}\,\text{yr}$, enormously exceeding both the Hubble time
    and any observation window.

  \item \textit{Level transitions}: peak strain $h_{0,\rm tr}^{\rm peak} \simeq
    6 \times 10^{-27}$ at $r = 100\,\text{au}$, with emission frequency $f_{\rm tr} \simeq
    833.03\,\text{MHz}$ for the $\{6,4,4\}\to\{5,4,4\}$ benchmark.  The signal
    lifetime $\tau_{\rm tr} \sim 10^3$--$10^6\,\text{yr}$ is long compared to any
    observation period but finite, giving a spectral width $\Delta f \sim
    \tau_{\rm tr}^{-1}/(2\pi) \ll 1\,\text{Hz}$. 

  \item The frequency-domain templates take analytic closed forms: an exponential-integral
    $E_1$ expression for the annihilation signal, and a complex Lorentzian shifted by
    the carrier frequency $\omega_{\rm ann}$ or $\omega_{\rm tr}$ for the two
    polarizations (Appendix~\ref{sec:templates}).  Both persistent signals trivially
    satisfy the steady-state cavity-response condition $\tau_{\rm signal} \gg
    \tau_{\rm ring} \sim 10^{-5}\,\text{s}$.
\end{itemize}

\paragraph{Binary-perturbed gravitational atom (Secs.~\ref{sec:binary-GA} and \ref{sec:pbh-rates}).}
Applying the Landau--Zener resonance formalism of Ref.~\cite{Kyriazis:2025fis} to
PBH binaries, we computed the characteristic strain of the transient GW burst produced
when the binary sweeps through the $\{2,1,1\}\to\{2,1,-1\}$ hyperfine transition.
The principal results are:
\begin{itemize}
  \item The peak GW frequency $f_p$ (Eq.~\eqref{eq:peak_freq}) lies naturally in the
    GHz band for host BH masses $M_{\rm BH} \sim 10^{-11}$--$10^{-12}\,M_\odot$ at $\alpha \approx0.14$, confirming that PBH binaries are the only astrophysical system that can populate wthe ADMX frequency range through binary-induced superradiance.

  \item The conservative duration-versus-response diagnostic
    $\Delta t \gg \tau_{\rm ring}$ is satisfied throughout the allowed $(\alpha,q)$
    parameter space of Fig.~\ref{fig:dwelltimeconstraint}, regardless of resonant
    frequency.  This is sufficient for applying the steady-state comparison within our conservative diagnostic model, but it is
    not a full detector-response calculation and is not by itself sufficient for detection.

  \item The GW characteristic strain at $r = 1\,\text{kpc}$ falls {\it fifteen
    orders of magnitude} below the ADMX threshold $h_{\rm min} \sim 10^{-22}$.
    Detection would require sources within to be unrealistically close.

  \item The early two-body PBH merger rate~\cite{Raidal:2024bmm} evaluated for the
    benchmark dichromatic mass function gives a characteristic event distance
    $d_{1\,\rm yr} \approx 1\,\text{kpc}$ for ADMX-relevant $(q = 10^{-4})$
    binaries.  Even with the maximally optimistic PBH clustering factor $\delta \sim
    10^5$, this reduces only to $\approx 22\,\text{pc}$—still approximately thirteen orders of
    magnitude in strain below the ADMX threshold.  We conclude that binary-driven
    level-transition events are not detectable with current ADMX sensitivity.
\end{itemize}

\subsection{Comparison to Related Work}

The GW signatures of superradiant clouds around {\it stellar-mass} BHs are well
studied~\cite{ArvanitakiDubovsky,Arvanitaki2015,Brito:2015,Isi:2019}, with constraints
on ultralight bosons from LIGO/Virgo spin measurements and non-detections of continuous
GWs establishing the most robust exclusion regions in the $(\mu, M_{\rm BH})$ plane
for $\mu \sim 10^{-13}$--$10^{-12}\,\text{eV}$~\cite{Ng:2019}.  The present
work extends this program into qualitatively new territory in two respects.

First, by focusing on PBH masses $M_{\rm BH} \lesssim 10^{-3}\,M_\odot$ we access
boson masses $\mu \sim 10^{-7}$--$10^{-5}\,\text{eV}$ for which no spin-measurement
constraints currently exist, and GW frequencies in the MHz--GHz range that are
inaccessible to LIGO-band searches.  High-frequency GW experiments such as ADMX and
the broadband detectors proposed in Refs.~\cite{AggarwalHF, Domcke:2022} offer the
only existing instrumental sensitivity in this band; our frequency-domain templates
(Appendix~\ref{sec:templates}) provide the waveform targets needed for a
matched-filter search.

Second, the binary-perturbation analysis builds directly on the Landau--Zener framework within~\cite{Baumann2021} and the recent extension
by Kyriazis \& Yang~\cite{Kyriazis:2025fis}, which derived the analytic GW strain
template for tidally driven transitions.  Previous applications of the Landau--Zener formalism
focused primarily on floating orbits, inspiral modifications,
dephasing of the binary GW signal, and the depletion of clouds in the context of LISA-band
sources~\cite{Baumann2021}, and on the resonance
history in Ref.~\cite{Tomaselli:2024}.  We are the first to apply the Kyriazis \& Yang
waveform template to PBH binaries in the MHz--GHz regime and to compare it directly
to the sensitivity of an existing high-frequency detector, yielding a quantitative
(negative) detectability result rather than a parametric estimate.

The bosenova bound on the cloud occupation number,
following Refs.~\cite{YoshinoKodama:2012, YoshinoKodama:2014}, is found to be
non-restrictive for QCD-axion-like couplings ($f_a \sim 10^{16}\,\text{GeV}$) in the
bulk of our PBH parameter space (Eq.~\eqref{eq:N_sat}), so the strain amplitudes we
derive are not suppressed by bosenova cycling.  For very weakly coupled bosons
($f_a \ll M_{\rm Pl}$) the cloud would cycle and the strain would be reduced; we note
this as a limitation and leave the quantitative treatment of bosenova-modified waveforms
to future work.

\subsection{Caveats and Limitations}

Several idealizations warrant discussion.

\paragraph{Spin distribution of PBHs.}

Our strain calculations assume that the host PBH initially lies above the relevant
Regge trajectory, and in several benchmark cases we take a near-extremal initial
spin in order to maximize the extractable rotational energy and hence the saturation
cloud occupation number.  The existence of an abundant population of PBHs with such
spins remains an open question.  Standard radiation-era formation from nearly Gaussian
perturbations tends to predict small natal spins~\cite{Chiba2017,DeLuca:2019pbhspin},
while early matter domination, highly nonspherical or non-Gaussian collapse,
phase-transition formation channels, subsequent PBH mergers, and environmental
accretion can produce larger spins in more model-dependent scenarios~\cite{Harada:2017pbhspin,Tichy:2008finalspin}.
Consequently, our amplitudes should be interpreted as optimistic conditional estimates
for the high-spin subpopulation, not as population-averaged predictions for all PBHs.
If only a fraction $f_{\rm spin}$ of PBHs of the relevant mass satisfy
$a_*>a_*^{\rm crit}(\alpha,m)$, the corresponding source counts and event rates
should be rescaled by this factor, or by the appropriate power of it when more than
one spinning PBH is required.  A full treatment of the joint PBH mass, abundance,
binary, and spin distributions is beyond the scope of this work and is left for
future study.

\paragraph{Isolated versus binary GAs.}
We have treated the isolated and binary channels independently, but in a realistic
PBH population both may be active simultaneously or sequentially.  Binary companions
that sweep through resonance before the isolated cloud reaches saturation will
partially deplete the cloud, suppressing the subsequent annihilation and level-transition
signals.  Conversely, an isolated cloud that has already
undergone significant annihilation before the binary resonance is crossded will produce
a weaker binary-induced burst than our estimates suggest.  A self-consistent treatment
that evolves both channels simultaneously remains to be carried out.

\paragraph{Detector response to chirps/transients.}

Our detectability estimates use the published ADMX-equivalent strain sensitivity as
a baseline and supplement it with a simple duration-versus-ring-up diagnostic.  This
is adequate for the persistent isolated-GA signals, but it is not a substitute for
a full detector-response analysis of binary-induced chirping/transient signals.  For such
sources, the relevant observable is the electromagnetic energy deposited in one or
more cavity modes after convolution with the detector transfer function, and this
response can differ from the naive expectation based solely on whether the source
duration is longer or shorter than $\tau_{\rm ring}$.  We have therefore interpreted
the binary-burst sensitivity estimates conservatively and regard a dedicated
matched-filter or deposited-energy analysis as an important next step.


\paragraph{Multi-level dynamics.}
Our binary waveform analysis follows Ref.~\cite{Kyriazis:2025fis} in treating the
$\{2,1,1\}\to\{2,1,-1\}$ transition as a two-level system.  In reality, multiple
superradiant levels may be populated simultaneously (Fig.~\ref{fig:Mass_fract}),
and the binary may sweep through several resonances in succession before merger.
The resonance history can be complex and depends on the level hierarchy and the
chirp timescale~\cite{Tomaselli:2024}; multi-level corrections to the waveform
template are beyond the scope of this paper.

\subsection{Outlook and Future Directions}

Despite the negative detectability conclusion for binary-induced bursts at current
ADMX sensitivity, several physically well-motivated avenues remain open.

\paragraph{Isolated-GA annihilation as a near-term target.}
The isolated-GA annihilation signal ($h_{0,\rm ann}^{\rm peak} \sim 10^{-21}$ at
145 au, $f_{\rm ann} \sim$\,MHz--GHz) is not subject to the event-rate limitations
that render binary-driven bursts undetectable.  For PBHs distributed throughout the
Milky Way dark-matter halo, sources within $\sim 200$\,au are plausibly present
with an abundance proportional to the local PBH dark-matter fraction $f_{\rm PBH}$.
The continuous, monochromatic nature of the signal ($\tau_{\rm ann} \gg t_{\rm Hubble}$,
$\Delta f \lesssim 10^{-33}\,\text{Hz}$) makes it an ideal target for coherent
integration strategies that accumulate signal-to-noise over extended observation
periods.  We identify this channel as the primary observational target within the
current ADMX band, and provide the explicit $E_1$-based frequency-domain template
(Appendix~A.1) needed to implement a matched-filter search.

\paragraph{Next-generation detector requirements for binary signals.}

For binary-induced bursts to become detectable, three improvements are needed in
combination: (i) enhancement in strain sensitivity
$h_{\rm min}$ (Eq.~\eqref{eq:ADMX_sensitivity}) through larger effective
volume, stronger magnetic field, reduced system temperature, or enhanced mode overlap;
(ii) broader frequency coverage, particularly at sub-GHz frequencies where heavier
and potentially more abundant PBHs contribute; and (iii) detector response and
readout strategies optimized for transient and frequency-drifting sources, rather
than only for steady-state monochromatic signals. Novel detector concepts---including
plasma haloscopes, LC circuit detectors, broadband bulk acoustic resonators, multimode
cavities, and correlated detector networks~\cite{AggarwalHF,Domcke:2022,Schenk:2025cavity,Blas:2026multimode,Amaral:2026gravnet}---may provide complementary
sensitivity in frequency ranges or signal classes not covered by current ADMX searches.

\paragraph{Constraints on the ultralight boson spectrum.}
Even without a detection, the non-observation of a persistent narrowband GW signal
at a given frequency places an upper bound on the product $f_{\rm PBH} \cdot
h_{0,\rm ann}^{\rm peak}$, which translates into a constraint on the PBH density
in the local dark-matter halo as a function of boson mass.  Deriving quantitative
spin-down constraints analogous to those obtained from LIGO-band continuous-wave
searches~\cite{Ng:2019}, but using ADMX non-detection data, is a direct
application of the templates presented in this paper and represents a novel intersection
of axion dark-matter searches with GW physics.

\paragraph{PBH spin distribution and formation constraints.}
The sensitivity of the GW strain to the initial BH spin (through $N_{\rm sat} \propto
a_* - a_*^{\rm crit}$) makes PBH-GA systems a probe of PBH formation mechanisms,
complementary to mass-function constraints from microlensing and CMB
observations~\cite{Carr:2020}.  A statistical population analysis combining the GW
signal distribution with the PBH spin distribution would allow simultaneous
constraints on both the PBH abundance and their formation spin, and is facilitated
by the analytic templates derived here.\\
\newpage


\noindent In sum, in this paper we have systematically studied the MHz--GHz gravitational-wave signatures of
superradiant boson clouds around primordial black holes.
Our results establish PBH--gravitational atom systems as among the
very few theoretically well-motivated sources of high-frequency gravitational waves,
and provide the complete set of waveform templates needed for targeted searches at
ADMX and successor experiments.  The gap between current sensitivity and the predicted
signal levels defines a concrete engineering target for next-generation high-frequency
GW detectors, and the boson mass range $\mu \sim 10^{-7}$--$10^{-5}\,\text{eV}$
probed by these signals remains largely unconstrained by any other observational
technique.

While isolated systems can produce long-lived, narrow-band signals with
potentially observable strain amplitudes under optimistic conditions,
binary-induced signals are generically suppressed both in amplitude and
event rate. Depending on binary parameters, some signals may even be transient compared to detector response times. This study highlights a clear experimental target:
improved strain sensitivity, broader frequency coverage, and detector response/readout
strategies optimized for transient and frequency-drifting signals are all essential
to probe this class of sources.Finally, the analytic
templates developed here provide concrete benchmarks for future high-frequency
gravitational-wave searches.

\section*{Acknowledgements}
We are grateful to Giovanni Maria Tomaselli and  Antonios Kyriazis for useful clarifications and discussions. LB, CE, and SP are supported in part by the U.S. Department of Energy, Office of Science, Office of High Energy Physics of U.S. Department of Energy under grant Contract Number DE-SC010107 (to SP).

\appendix

\section{Frequency Domain Strain Templates}\label{sec:templates}

The detectability of a gravitational-wave signal is most cleanly expressed in terms of the characteristic strain, which is defined from the Fourier transform of the metric perturbation as
$h_c(f) = 2 f \lvert \tilde h(f) \rvert$.
While the raw Fourier amplitude $\tilde h(f)$ depends on the details of the observation, such as the total duration of the time series and the specific sampling, the combination $h_c(f)$ isolates the physically relevant quantity: the accumulated signal power at a given frequency, including the contribution from the number of coherent wave cycles. For a quasi-monochromatic source observed over an observation time $T$, one has approximately
$h_c(f_0) \sim h_0 \sqrt{f_0 T}$,
so that the characteristic strain increases with the square root of the effective number of cycles. This makes $h_c(f)$ the appropriate quantity for comparing theoretical signals to the sensitivity of gravitational-wave detectors.

A detector is characterized by its one-sided noise power spectral density $S_n(f)$. The corresponding noise amplitude is given by
$h_n(f) = \sqrt{f S_n(f)}$.
Signal power and detector noise are compared at the same level: both $h_c(f)$ and $h_n(f)$ have the same dimensions and represent amplitudes per logarithmic frequency interval. A source is detectable when its characteristic strain exceeds the noise curve, and the expected signal-to-noise ratio is
\begin{equation}
SNR^2 = \int_0^{\infty} \frac{h_c^2(f)}{f^2 S_n(f)} \, df = \int_0^{\infty} \frac{h_c^2(f)}{h^2_n(f)} \, d(\ln f).
\end{equation}
This expression shows explicitly why $h_c(f)$ is the correct quantity to compare with detector sensitivity. Any direct comparison using $\tilde h(f)$ alone would be misleading, because $\tilde h(f)$ scales with the chosen Fourier window and does not represent a physical amplitude.

In the context of black hole superradiance and axion annihilation or level transitions, the characteristic strain provides a unified language for evaluating detectability against detector strain curves. The signals are intrinsically narrow-band and can be long-lived, so $h_c(f)$ reflects both the instantaneous gravitational-wave amplitude and the coherence of the emission process. Even if the raw strain is extremely small, the large number of coherent cycles at the transition or annihilation frequency can raise $h_c(f)$ above detector noise. Once expressed in terms of $h_c(f)$, the comparison to detector sensitivity curves is immediate: gravitational-wave observatories are sensitive to signals whenever $h_c(f)$ lies above (or not far below) $h_n(f)$ at the relevant frequency. This is the fundamental criterion for detection prospects.

\subsection{Annihilation Strain} 
The bulk of this paper has shown that there are multiple potentially detectable signals in a gravitational atom system. These signals can arise from both an isolated gravitational atom or due to the presence of a binary companion. In the first case, we have shown that detectable signals arise from two separate phenomenon: level transitions and boson annihilations. 

The first frequency domain strain that we consider arises due to the axion annihilation within a specified level. Again, the time domain strain is specified in Eq \eqref{eq:h0ann_env} and the only time dependence arises from the decay of the state population in Eq. \eqref{eq:N_ann}: 

\begin{equation}
 h_{0,\text{ann}}(t) = \frac{N_{\max}}{1 + \Gamma_a N_{\max} t}  \sqrt{\frac{4 G_N}{r^2 \omega_{\text{ann}}} \Gamma_a} 
\end{equation}
where again we have made the assumptions that $\alpha$ and the superradiance rates are unchanging in time to obtain an approximate form of the annihilation strain amplitude. Again, this allows us to take a fourier transform to obtain an analytic form of the annihilation strain as a function of frequency. 
We begin with the one-sided Fourier transform,

\begin{equation}
\begin{aligned}
    \tilde{h} _{0,\text{ann}}(\omega)
    &= \frac{1}{\sqrt{2\pi}} 
       \int_0^{\infty} \frac{A\, e^{-i\omega t}}{1 + \beta t}\, dt,
       \qquad \beta > 0, \\[0.5em]
\end{aligned}
\end{equation}
where we will consistently use the convention of $f = \frac{\omega}{2 \pi}$.
\begin{equation}
\begin{aligned}
    A = N_{\max}&\,\sqrt{\frac{4 G_N\,\Gamma_a}{r^2 \omega_{\text{ann}}}}, \hspace{3em} \beta = \Gamma_a N_{max}.
\end{aligned}
\end{equation}
Let \(u = 1 + \beta t\), so that \(t = (u - 1)/\beta\) and \(dt = du / \beta\).
Substituting, we find
\begin{equation}
    \tilde{h} _{0,\text{ann}}
    = \frac{A}{\sqrt{2\pi}\, \beta}
      e^{\,i\omega / \beta}
      \int_1^{\infty} \frac{e^{-i(\omega / \beta) u}}{u}\, du.
    \label{eq:h_int_form}
\end{equation}

To evaluate the integral, we introduce a causal regulator \(+i\epsilon\) to ensure convergence
and define the exponential integral \(E_1(x)\) in its equivalent forms:
\begin{equation}
    E_1(x)
    \equiv
    \int_1^{\infty} \frac{e^{-t x}}{t}\, dt
    =
    \int_x^{\infty} \frac{e^{-u}}{u}\, du,
    \qquad (|\arg x| < \pi).
    \label{eq:E1_def}
\end{equation}
Comparing with Eq.~\eqref{eq:h_int_form}, we immediately identify
\[
    x = i\frac{\omega}{\beta} + i \epsilon,
\]
where the infinitesimal imaginary term \(+i \epsilon\) enforces the causal (\(+i\epsilon\)) prescription.
Thus,
\begin{equation}
    \int_1^{\infty} \frac{e^{-i(\omega/\beta)u}}{u}\, du
    = E_1\!\left(i\frac{\omega}{\beta} + i \epsilon \right).
    \label{eq:E1_relation}
\end{equation}
Substituting Eq.~\eqref{eq:E1_relation} into Eq.~\eqref{eq:h_int_form}, we obtain the compact result
\begin{equation}
   \tilde{h} _{0,\text{ann}}
    = \frac{A}{\sqrt{2\pi}\, \beta}
      e^{\,i\omega / \beta}
      E_1\!\left(i\frac{\omega}{\beta} + i\epsilon\right),
    \label{eq:h_E1_result}
\end{equation}
where \(i\epsilon\) denotes the same causal regulator used in the definition of \(E_1\). To rewrite Eq.~\eqref{eq:h_E1_result} in terms of the exponential integral $\mathrm{Ei}$ and more familiar functions, we use the relation
\begin{equation}
E_1(z) = -\mathrm{Ei}(-z),
\label{eq:E1-Ei}
\end{equation}
and continue it to complex arguments along the causal branch.
For imaginary arguments, the continuation gives
\begin{equation}
\mathrm{Ei}(ix + i\epsilon)
= \mathrm{PV}\big[\mathrm{Ei}(ix)\big]
 + i\pi\mathrm{Sign}(x),
\label{eq:Ei-branch}
\end{equation}
where $\mathrm{PV}\big[\mathrm{Ei}(ix)\big] $ refers to the principal value of the exponential integral and the second term represents the usual $i\pi$ phase picked up when crossing the branch.
Similarly, the logarithm acquires the standard phase shift across its branch cut,
\begin{equation}
\log(i\omega) - \log(-i\omega) = i\pi\mathrm{Sign}(\omega),
\label{eq:log-jump}
\end{equation}
corresponding to the same causal choice of branch.

\begin{figure}[t!]
\begin{center}
\includegraphics[width = 0.9\textwidth, height = 0.5\textwidth ]{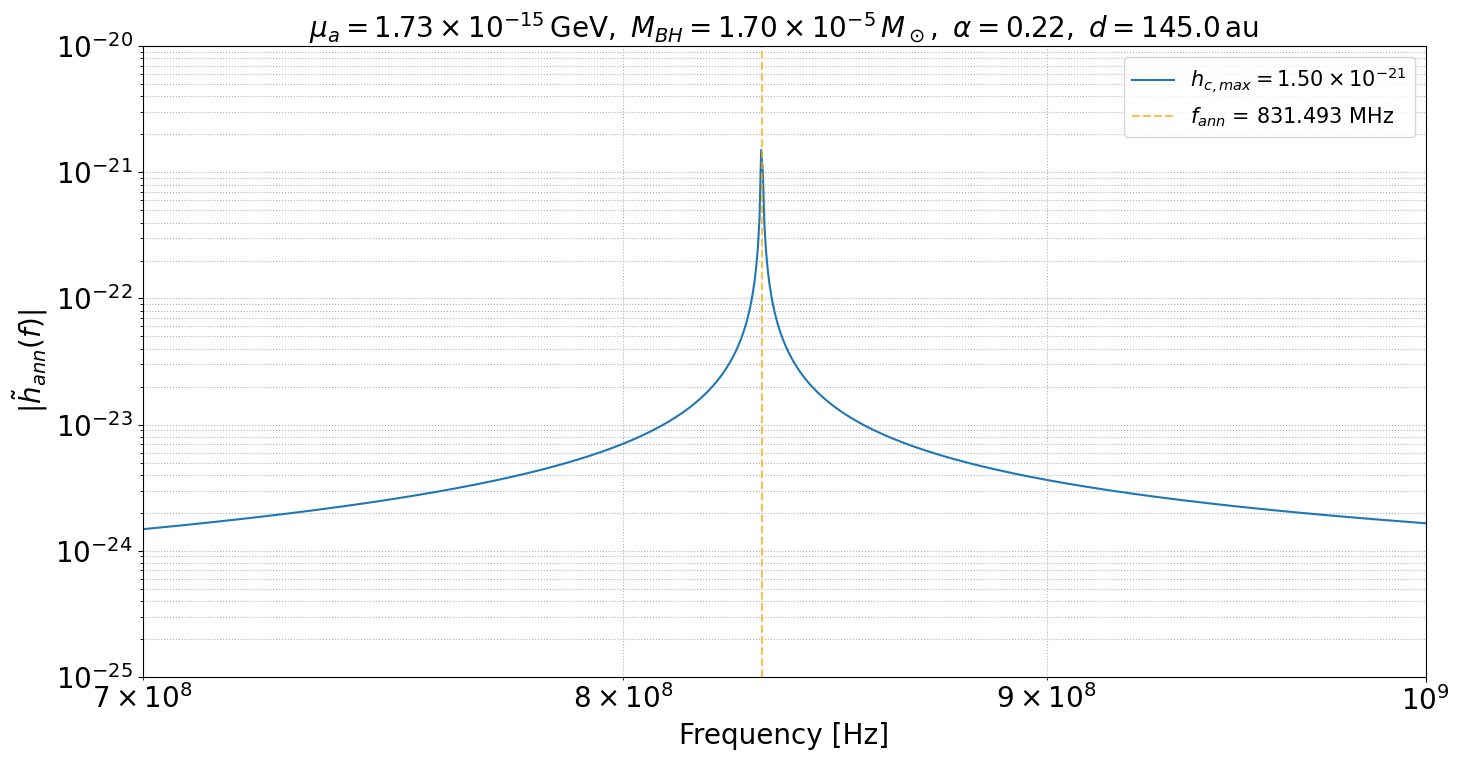}
\caption{Characteristic strain  of the Fourier transform for the annihilation strain. Input parameters match those from Fig. \ref{fig:annihilation_strain} for the $n =2 , l = m =1$ state to obtain the strongest annihilation strain amplitude. The peak strain matches those expected in Ref \cite{yqms-lznb}.}
\label{fig:anihhilation_characteristic_strain}
\end{center}
\end{figure}

Using Equations \eqref{eq:E1-Ei}-\eqref{eq:log-jump}, we can rewrite equation \eqref{eq:h_E1_result} as:
\begin{align}
 \tilde{h}_{0,\text{ann}}
  &= \frac{A}{\sqrt{2\pi}\,\beta}\,e^{\,i\omega/\beta}
     \left[
       -\,\mathrm{Ei}\!\left(-\frac{i\omega}{\beta}\right)
       - i\pi\,\mathrm{sgn}(\omega)
     \right].
  \label{eq:FT-corrected-logs}
\end{align}

Finally, restoring $A$ and $\beta=\Gamma_a N_{\max}$ gives
\begin{equation}
\tilde{h}_{0,\text{ann}}(\omega)=
\frac{1}{\sqrt{2\pi}}\sqrt{\frac{4G_N}{r^2\omega_{\rm ann}\Gamma_a}}\;
e^{\,i\omega/(\Gamma_a N_{\max})}\,
\left[
-\,\mathrm{Ei}\!\left(-\frac{i\omega}{\Gamma_a N_{\max}}\right)
- i\pi\,\mathrm{sgn}(\omega)
\right].
\label{eq:final}
\end{equation}
which is identical to \eqref{eq:FT-corrected-logs} upon using \eqref{eq:log-jump} and restoring all constants.

In order to return the oscillatory term to our frequency domain template, we need to multiply the envelope by \(e^{\pm i\omega_{\rm ann} t}\). This has the effect of shifting our frequency, so we can describe the Fourier transform of our envelope as:

\begin{equation}
\mathcal{F}\!\left[h_0(t)\,e^{\pm i\omega_{\rm ann} t}\right]
= \tilde h\!\left(\omega \mp \omega_{\rm ann}\right)
\label{eq:freq_shift}
\end{equation}

With
\(\cos = \tfrac12(e^{i\phi_0}e^{i\omega_{\rm ann} t}+e^{-i\phi_0}e^{-i\omega_{\rm ann} t})\),
\(\sin = \tfrac1{2i}(e^{i\phi_0}e^{i\omega_{\rm ann} t}-e^{-i\phi_0}e^{-i\omega_{\rm ann} t})\),
the polarization spectra are
\begin{align}
\tilde h_+(\omega)
&=\frac{1+\cos^2(\iota)}{4}\
\Big[e^{i\phi_0}\,\tilde h_(\omega-\omega_{\rm ann})
+e^{-i\phi_0}\,\tilde h(\omega+\omega_{\rm ann})\Big],
\label{eq:Hplus_freq}\\[2pt]
\tilde h_\times(\omega)
&=\frac{\cos(\iota)}{2i}
\Big[e^{i\phi_0}\,\tilde h(\omega-\omega_{\rm ann})
-e^{-i\phi_0}\,\tilde h(\omega+\omega_{\rm ann})\Big].
\label{eq:Hcross_freq}
\end{align}
Substituting $\tilde{h} $ gives the explicit closed forms:
\begin{align}
\tilde h_+(\omega)
&=\,\frac{N_{\max}}{\sqrt{2\pi}}\sqrt{\frac{4G_N}{r^2\omega_{\rm ann}}} \frac{1+\cos^2(\iota)}{4}
\!\left[
e^{i\phi_0}\,e^{\,i(\omega-\omega_{\rm ann})/\beta}
\Big(-2\,\mathrm{Ei}(\!\tfrac{i(\omega-\omega_{\rm ann})}{\beta})
+ i\pi\,\mathrm{Sign}(\omega-\omega_{\rm ann})\Big)
\right.\nonumber\\[-2pt]
&\hspace{6.6em}\left.+
e^{-i\phi_0}\,e^{\,i(\omega+\omega_{\rm ann})/\beta}
\Big(-2\,\mathrm{Ei}(\!\tfrac{i(\omega+\omega_{\rm ann})}{\beta})
+ i\pi\,\mathrm{Sign}(\omega+\omega_{\rm ann})\Big)
\right],
\label{eq:Hplus_explicit}\\[3pt]
\tilde h_\times(\omega)
&=\frac{N_{\max}}{\sqrt{2\pi}}\sqrt{\frac{4G_N}{r^2\omega_{\rm ann}}} \frac{\cos(\iota)}{2i}
\!\left[
e^{i\phi_0}\,e^{\,i(\omega-\omega_{\rm ann})/\beta}
\Big(-2\,\mathrm{Ei}(\!\tfrac{i(\omega-\omega_{\rm ann})}{\beta})
+ i\pi\,\mathrm{Sign}(\omega-\omega_{\rm ann})\Big)
\right.\nonumber\\[-2pt]
&\hspace{6.8em}\left.-
e^{-i\phi_0}\,e^{\,i(\omega+\omega_{\rm ann})/\beta}
\Big(-2\,\mathrm{Ei}(\!\tfrac{i(\omega+\omega_{\rm ann})}{\beta})
+ i\pi\,\mathrm{Sign}(\omega+\omega_{\rm ann})\Big)
\right].
\label{eq:Hcross_explicit}
\end{align}

\subsection{Isolated Level Transition Strain}\label{sec: isolated template}
We next develop a frequency domain template of the transition amplitude gravitational strain outlined in Eq. \eqref{eq:transition_envelope} by taking a Fourier transform of the approximate solution. Since both the states grow exponentially and approximately independently for the majority of the lifetime of the excited state, we will be approximating the form of both the ground and excited states as $\frac{dN_g}{dt} = \Gamma^{sr}_g N_g + \Gamma_tN_gN_e$ and $\frac{dN_e}{dt} = \Gamma^{sr}_e N_e -\Gamma_tN_gN_e$. The resulting time dependence of each state is then an exponential of the form $ N_g(t)= N_0 \exp{(\Gamma_g^{SR} + \Gamma_t N_e )t}$ and $ N_e(t)= N_0 \exp{(\Gamma_e^{SR} - \Gamma_t N_g )t}$. The transition amplitude as a function of time, assuming a constant in time superradiant transition rate $\Gamma_t$, gives us a transition strain of the form 
\begin{equation}
    h_{0,tr} \approx \sqrt{\frac{4G_N}{r^2 \omega_{tr}} \Gamma_t N^g_{0} N_0^e }  \exp{[ \frac{1}{2} (\Gamma^{sr}_e + \Gamma^{sr}_g + \Gamma_t N_e - \Gamma_t N_g )t ] }
\end{equation}
Under these approximations we can take a Fourier transform to find the functional form of our strain. Doing so, we obtain the following expression: 
\begin{equation}
h_{0,tr}(\omega) =\frac{1}{2\pi}
\frac{\sqrt{\frac{4 G_N}{r^2 \omega_{\mathrm{tr}}} \, \Gamma_tN_0^gN_0^e}}
{ \; \left( \Gamma^{sr}_e + \Gamma^{sr}_g + \Gamma_t N_g - \Gamma_t N_e    - i \omega \right) } \bigg( - 1 +{[ \frac{1}{2} (\Gamma^{sr}_e + \Gamma^{sr}_g + \Gamma_t N_e - \Gamma_t N_g - i \omega)T]} \bigg)
\end{equation}

with $T$ being the time taken from the start of superradiance to when the excited state has transitioned completely. This is the approximate form of our gravitational wave signal in the frequency domain due to level transitions within the gravitational atom. Since detectors look at the frequency decomposition of these signals, this is ultimately the desired form for the gravitational strain. 
Again, to include the physical oscillations of the gravitational wave and its orientation,
we can treat the transition signal as an amplitude (or envelope) modulated by a carrier
frequency corresponding to the transition energy. In other words, the total strain is just
our smooth envelope multiplied by an oscillatory factor at $\omega_{\rm tr}$:
\[
h_{\rm tr}(t)=h_{0,\rm tr}(t)\cos(\omega_{\rm tr} t+\phi_0)
=\tfrac{1}{2}\,h_{0,\rm tr}(t)
\left(e^{i(\omega_{\rm tr} t+\phi_0)}+e^{-i(\omega_{\rm tr} t+\phi_0)}\right).
\]

When we Fourier transform this, the oscillatory pieces simply shift the spectrum to
$\pm\omega_{\rm tr}$, giving
\begin{equation}
\mathcal{F}\!\left[h^{\text{env}}_{\rm tr}(t)e^{\pm i\omega_{\rm tr} t}\right]
=\tilde{h}^{\text{env}}_{\rm tr}(\omega\mp\omega_{\rm tr}).
\label{eq:tr_shift}
\end{equation}
So the frequency-domain strain becomes two terms centered on the transition frequency,
each weighted by the phase $\phi_0$:
\begin{equation}
\tilde{h}_{\rm tr}(\omega)
=\frac{1}{2}\left[
e^{i\phi_0}\,\tilde{h}_{0,\rm tr}(\omega-\omega_{\rm tr})
+e^{-i\phi_0}\,\tilde{h}_{0,\rm tr}(\omega+\omega_{\rm tr})
\right].
\label{eq:tr_cos_shift}
\end{equation}

Finally, we can include the detector-frame polarization dependence. The observed strain
depends on the inclination angle $\iota$ of the system relative to the line of sight.
The two standard polarizations then take the form
\begin{align}
\tilde{h}^{(+)}_{\rm tr}(\omega)
&=\frac{1+\cos^2\!\iota}{4}
\Big[
e^{i\phi_0}\,\tilde{h}^{\text{env}}_{\rm tr}(\omega-\omega_{\rm tr})
+e^{-i\phi_0}\,\tilde{h}^{\text{env}}_{\rm tr}(\omega+\omega_{\rm tr})
\Big],
\label{eq:Hplus_tr}\\[4pt]
\tilde{h}^{(\times)}_{\rm tr}(\omega)
&=\frac{\cos\!\iota}{2i}
\Big[
e^{i\phi_0}\,\tilde{h}^{\text{env}}_{\rm tr}(\omega-\omega_{\rm tr})
-e^{-i\phi_0}\,\tilde{h}^{\text{env}}_{\rm tr}(\omega+\omega_{\rm tr})
\Big].
\label{eq:Hcross_tr}
\end{align}
These two polarizations represent the characteristic “plus” and “cross” patterns of the
transition signal, differing only by a relative phase and the inclination weighting.
In practice, both appear as narrow peaks in the frequency spectrum at $\omega_{\rm tr}$,
modulated by the growth and decay encoded in $\tilde{h}^{\text{env}}_{\rm tr}(\omega)$.

 The spectral width $\Delta \omega$ of the template is
controlled entirely by the characteristic timescale over which the envelope
$A(t)$ varies (i.e.\ by the relevant superradiant and transition rates),
\begin{equation}
    \Delta \omega \sim \Gamma_{\mathrm{tot}} \sim \mathcal{O}\left(\text{yr}^{-1}\right)
    \;\;\Rightarrow\;\;
    \Delta f \sim \frac{\Delta \omega}{2\pi} \ll 1~\mathrm{Hz},
\end{equation}
even though the line is centered at $f_{\mathrm{tr}} = \omega_{\mathrm{tr}}/2\pi
\sim \mathrm{MHz}$. Consequently, the frequency-domain signal appears as an
extremely narrow (sub-Hz) Lorentzian line located at the MHz transition
frequency. In general, the carrier frequency fixes the centering of the line position, while the slow
population evolution ($\mathcal{O}(10^3)$ years) sets the linewidth. Faster population evolutions will therefore result in Lorentzian signals with much wider frequency resolution, whereas slowly evolving states will appear as increasingly narrow, near-monochromatic lines in the frequency spectrum.

\begin{figure}[h!]
    \centering
    \makebox[\textwidth][c]{\includegraphics[width=0.7\linewidth]{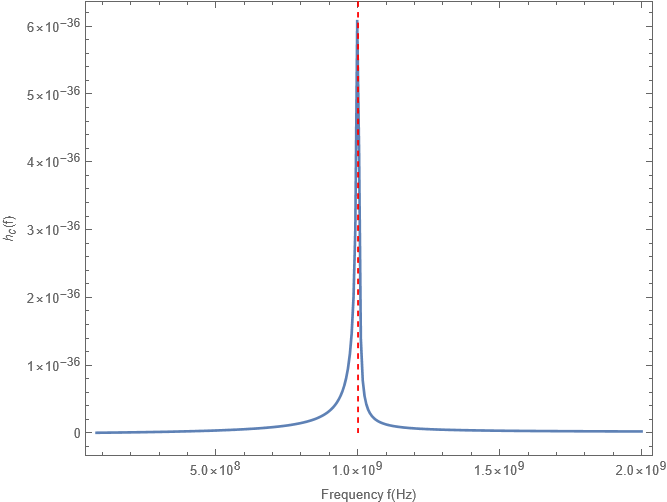}}
    \caption{Computation of the binary perturbation characteristic strain $h_c(f) = 2f|\tilde{h}(f)|$ with $\tilde{h}(f)$
given by Eq.~\eqref{eq:binary_h_plus_fourier}. Benchmark parameters from Sec.~\ref{sec:binary-GA} are used to match the high frequency detection range of resonant cavity based detectors, such as ADMX. The above plot shows the characteristic strain as a function of frequency for the benchmark binary GA system, with the the peak frequency $f_p = 1.001\,\text{GHz}$ visualized by the red dashed line.}
    \label{fig:pbh_charateristic}
\end{figure}

\subsection{Binary Level Transition Strain}\label{sec:binary template}

The frequency domain template for the “plus” polarization for binary level transition strain given within Eq. \eqref{eq:binary_signal_h+} is already defined by Eq. \eqref{eq:binary_h_plus_fourier}, where a detailed derivation using the stationary phase approximation can be found within \cite{Kyriazis:2025fis}. The frequency domain template for the “cross” polarization can also be easily obtained using the same method: Rewriting Eq. \eqref{eq:binary_signal_hx} into the form of:

\begin{equation}
    h_{\times, 211}(t) = h_0 \cos\iota \; \left[\frac{e^{-2i\Delta m\varphi(t_{\rm re})}Q(t_{\rm re}) - e^{2i\Delta m\varphi(t_{\rm re})}Q^*(t_{\rm re})}{2i}\right]
\end{equation}

\noindent by expanding the imaginary component, then plugging into the Fourier transform:

\begin{equation}
    \tilde{h}_{+,\times}(\omega) = \int dt\, h_{+,\times}(t)\, e^{i\omega t}.
\end{equation}
we obtain the following expression:

\begin{equation}
    \tilde{h}_{\times}(\omega) = \frac{h_0 \cos\iota} {2i} e^{i\omega r} \int dt \left[ e^{i\omega t - 2i\Delta m\varphi(t)} Q(t) - e^{i\omega t + 2i\Delta m\varphi(t)} Q^*(t) \right]
\end{equation}
where we renamed $t_{\rm re} \to t$.

Evaluating the integral through the same method of stationary phase approximation within \cite{Kyriazis:2025fis} with the same arguments so that the first integral vanishes, we obtain: 

\begin{equation}
    \tilde{h}_{\times}(\omega) = -\frac{\sqrt{\pi}h_0 \cos \iota}{2i\sqrt{\gamma|\Delta m|}} Q^* (t_{+}(\omega))e^{i\Psi_{+}(\omega)}
\end{equation}
where $t_{+}(\omega) = \frac{\omega - \omega_c}{2|\Delta m|\gamma}$ donates the non-zero stationary point and the phase argument $\Psi_{+}(\omega)$ is given by $\Psi_{+}(\omega) = \omega r + \frac{(\omega - \omega_{c})^{2}}{4|\Delta m|\gamma} - \frac{\pi}{4}$, identical to the form we defined within Eq. \eqref{eq:binary_h_plus_fourier}. Substituting in $Q(t)$ as detailed within \cite{Kyriazis:2025fis}, we obtain:

\begin{equation}
    \tilde{h}_{\times}(f) = -\frac{2}{i}h_0\cos\iota\,\sqrt{\pi}\,|\Delta m|^2\,i\,
    e^{i\Psi_+(f)}\,
    \frac{\sqrt{z}}{|\Gamma| - i\pi(f-f_c)}\,
    e^{-\pi z}\,
    \exp\!\left[-2z\arctan\!\left(\frac{\pi(f-f_c)}{|\Gamma|}\right)\right]\,
\end{equation}
where we rewrote the frequency as $f = \omega / 2\pi$. Comparing to Eq. \eqref{eq:binary_h_plus_fourier}:

\begin{equation*}
      \tilde{h}_+(f) = h_0(1+\cos^2\iota)\,\sqrt{\pi}\,|\Delta m|^2\,i\,
    e^{i\Psi_+(f)}\,
    \frac{\sqrt{z}}{|\Gamma| - i\pi(f-f_c)}\,
    e^{-\pi z}\,
    \exp\!\left[-2z\arctan\!\left(\frac{\pi(f-f_c)}{|\Gamma|}\right)\right]\,
\end{equation*}
we notice that similar to isolated level transitions described within Sec.~\ref{sec: isolated template}, the two polarizations differ only by the inclination weighting and a relative phase of $\pi/2$ given by the factor of $-2/i$. Figure \ref{fig:pbh_charateristic} showcases  the characteristic strain for the benchmark parameters that produce UHF-GWs with frequency $\sim$1 GHz.
\bibliographystyle{JHEP}

\normalem
\bibliography{biblio.bib}
\end{document}